\documentclass[10pt,journal,compsoc]{IEEEtran}

\usepackage{float}
\usepackage{ifpdf}
\usepackage{multirow}
\usepackage{siunitx}
\usepackage{booktabs}
\usepackage{blindtext}
\usepackage{hyperref}
\usepackage{upquote}
\usepackage{array}
\usepackage{amssymb, amsfonts}
\usepackage[table]{xcolor}
\usepackage[ruled, vlined, linesnumbered]{algorithm2e}
\usepackage{graphicx}
\usepackage{textcomp}
\usepackage{caption}
\usepackage{comment}
\usepackage{adjustbox}

\usepackage{mathtools}
\usepackage{mathrsfs}
\usepackage{commath}
\usepackage{amsmath}
\usepackage{nicefrac}
\usepackage{cite}
\usepackage[inline]{enumitem}
\usepackage[switch]{lineno}

\newcommand{\comm}[1]{}

\def\BibTeX{{\rm B\kern-.05em{\sc i\kern-.025em b}\kern-.08em
    T\kern-.1667em\lower.7ex\hbox{E}\kern-.125emX}}

\begin{document}

\title{AI-enabled Sound Pattern Recognition on Asthma Medication Adherence: Evaluation with the RDA Benchmark Suite}

\author{
\IEEEauthorblockN{
Nikos Dimitris Fakotakis,
Stavros Nousias,
Gerasimos Arvanitis,
Evangelia I. Zacharaki and \\
Konstantinos Moustakas \\}
\vspace{1em}
\IEEEauthorblockA{$^1$ \footnotesize Department of Electrical and Computer Engineering, University of Patras, Greece\\
}


}

\IEEEtitleabstractindextext{%
\begin{abstract}
Asthma is a common, usually long-term respiratory disease with negative impact on global society and economy. Treatment involves using medical devices (inhalers) that distribute medication to the airways and its efficiency depends on the precision of the inhalation technique. There is a clinical need for objective methods to assess the inhalation technique, during clinical consultation. Integrated health monitoring systems, equipped with sensors, enable the recognition of drug actuation, embedded with sound signal detection, analysis and identification, from intelligent structures, that could provide powerful tools for reliable content management. Health monitoring systems equipped with sensors, embedded with sound signal detection, enable the recognition of drug actuation and could be used for effective audio content analysis. This paper revisits sound pattern recognition with machine learning techniques for asthma medication adherence assessment and presents the \textit{Respiratory and Drug Actuation (RDA)} Suite \footnote{https://gitlab.com/vvr/monitoring-medication-adherence/rda-benchmark} for benchmarking and further research. The RDA Suite includes a set of tools for audio processing, feature extraction and classification procedures and is provided along with a dataset, consisting of respiratory and drug actuation sounds. The classification models in RDA are implemented based on conventional and advanced machine learning and deep networks' architectures. This study provides a comparative evaluation of the implemented approaches, examines potential improvements and discusses on challenges and future tendencies.
\end{abstract}

}

\maketitle

\IEEEdisplaynontitleabstractindextext

%
\IEEEpeerreviewmaketitle

\section{Introduction}
\label{section:introduction}

Asthma is a chronic inflammatory condition of the airways, with over 235 million people \cite{WHO1, WHO2, WHO3} suffering worldwide. The direct cost associated with asthma has risen a significant burden on society and healthcare systems \cite{kopka8, kopka9}. Asthma is a common disease among children and one of the most common chronic conditions. It is characterized by recurrent attacks of wheezing, breathlessness, dyspnea, chest tightness and coughing, known as asthma attacks \cite{kaufman2009strategies}. It deteriorates the quality of life for patients and their families up to a level, leading to increased healthcare cost, poor clinical outcomes and increased morbidity rates \cite{sears1990regular}. The variety of obstructive respiratory diseases \cite{asher1, hornclark} reveals the importance for innovative models that could help patients face their unpleasant condition and enjoy a better quality of life \cite{akinbami2016changing}. There is an evidence that adherence to asthma treatment is variable, hindering the proper management of the disease \cite{british1982death}. Detailed constructive feedback from clinicians on inhaler usage may motivate patients to focus more on treatment adherence, which in turn may improve their quality of life, prevent exacerbation and hospitalization events and, eventually, reduce mortality rates associated with chronic respiratory diseases \cite{krishnan2004corticosteroid}.

Inhaled aerosol therapies are the main treatment of obstructive lung diseases \cite{boutayeb2005burden}. Inhaler based monitoring devices were introduced at the beginning of the 1980s and, since then, have been developed, mainly, for the proper assessment of medication adherence. These devices are presented in section \ref{section:hardware}. Aerosol devices deliver a fixed medication dose, rapidly and directly into the airways, from a pressurized canister containing a medication/propellant mixture \cite{fuller1995diskustm, malmberg2010inspiratory}. The efficient and effective management of asthma is strongly connected with patient adherence to the prescribed action plan, while reduced adherence has been strongly linked with significant indicators of health degradation. Active feedback may encourage patients to improve their adherence and manage their condition, better. For this purpose, specialists have developed methodologies to monitor inhaler users and understand if patients use their inhaler devices, with the appropriate technique and at the correct time duration. Audio process phenomenology requires the transformation of the measured data to extract the desired information (section \ref{section:featureextraction}). The current study presents machine learning and, mainly, deep learning approaches, in order for the researchers, to develop classification models that given as input the monitored acoustic signals, automatically recognize the phase of respiration and detect the drug inhalation onset (section \ref{section:classification}). Often, time-frequency analysis of the respiratory signals \cite{cavusoglu2007efficient, sa2002automated, wilson1982algorithms} is performed prior to classification, to extract a variety of features, critical for extensive analysis. Acoustic analysis of breathing has been employed to detect the different phases of respiration, such as inhalation, exhalation and drug actuation \cite{hossain2002respiratory, huq2012acoustic, yadollahi2006robust}. Many of these works are related to personalized management services on obstructive respiratory diseases, aiming to provide methodologies for medication adherence monitoring procedures \cite{aggarwal2014use}. Specifically, these studies either focus on the deployment of device-integrated solutions, using pressure-activated switches \cite{howard2014electronic, pilcher2016validation} or intelligent systems on ambient sound analysis \cite{holmes2014acoustic}. Despite their coalescent differences, every approach targets on the detection of drug actuation and recognition of respiratory events, for the right drug assessment and procedure management.

Signal processing techniques can effectively extract useful information. Audio processing problems can lead to some complex and intricate approaches, for performing pattern extraction of critical information, especially on noisy and sometimes incomplete (i.e., time series with missing values) sound measurements \cite{hartmann2004signals, raichel2006science}. However, they can be quite difficult to be developed. The adherence of patients to their medication intake, in terms of prescribed dosage and careful usage of inhaler devices, is critical for controlling the disease, as 24\% of asthma exacerbation and 60\% of hospitalization are caused by poor medication adherence \cite{baarnes2015asthma}. Studies suggest that up to 67\% of clinicians cannot describe the steps correctly or demonstrate correct inhaler usage, so we focus on the optimization of adherence and on the management of non-adherence, through the usage of systems with methodologies that consider patient's preferences, on the treatment and care decisions \cite{arcy25, pritch}. For these experiments, it was used a pressurized metered-dose inhaler (pMDI), where patients were instructed to actuate the canister of the pMDI, as they begin a "slow" and "deep" inhalation \cite{chrystyn2009not}. To ensure that the medication reaches the lower airways, the inhalation in drug actuation must be steadily below 90 L/min \cite{al2007can, ammari2013optimizing}. Because the majority of patients perform at least one step of the inhalation technique incorrectly (insufficient respiratory effort), systematic training is required to achieve optimal inhaler technique \cite{sulaiman2017objective}. Through breathing (inhalation and exhalation) the respiratory system facilitates the exchange of gases, between the air and the blood and between the blood and the body's cells \cite{sims2011aerosol}. Modern signal analysis techniques have been applied to extract features from inhalation sounds that characterize the events. Hence, we hypothesized that by analyzing the acoustics of inhalation, in a group of patients with a variety of respiratory and non-respiratory diseases, we could assess the accuracy, the sensitivity and the specificity of the models, related with inhalation sound recognition \cite{seheult2014acoustic}.

\section{Methodology}

This paper presents an extensive review and discussion on the state-of-the-art methods and tools for acoustic analysis and content-based audio classification of inhaler sounds on medication adherence, which could be used to improve the techniques on aerosol therapy \cite{crocker1997encyclopedia}. We try to cover a large part of the existing material, so all points of interest need to be included, to capture from one corner of the topic to the current status of the research and make the research of broad interest, but focusing only on inhaler's and respiratory sounds.

The methodology that we followed aimed to include works utilizing pMDI inhalers for audio analysis and recognition purposes. Our research includes articles that use classical algorithms and machine learning approaches for acoustic signal analysis, detection and recognition. The state of the art begins with methods from 2010, which mainly use decision trees as a technique for the identification of the signal and continues with supervised learning methods, aiming to classify respiratory sounds obtained from pMDIs. This scientific sub-field is referred as "sound analysis, detection and recognition" and the search query was formed as follows: (("inhalers' sounds") AND ("identification" OR "recognition" OR "classification") AND ("machine learning" OR "deep learning")). We retrieved papers published from January 1, 2010 until December 31, 2021. In addition to the articles extracted from this search, we also examined works published in the same time range using manual search. The main findings of the algorithms detailed in this review, suggest that temporal and spectral audio-based features of inhaler sounds can be used to assess the inhalation techniques, objectively.

Upon the literature review, we developed the RDA Benchmark suite, in which we implemented techniques from original research articles and, also, used it for a comparative analysis. The literature search has identified boosting approaches, decision trees' logic and deep neural networks for respiratory and drug actuation classification. The models are built in a stage-wise fashion and arbitrary differentiable loss functions are introduced for improvement of algorithms' performance and to encounter for possible over-fitting. We deployed the several methods and evaluated them on the RDA dataset, which includes the actuation, inhalation and exhalation classes.

\begin{figure*}
    \centering
    \includegraphics[width=\linewidth]{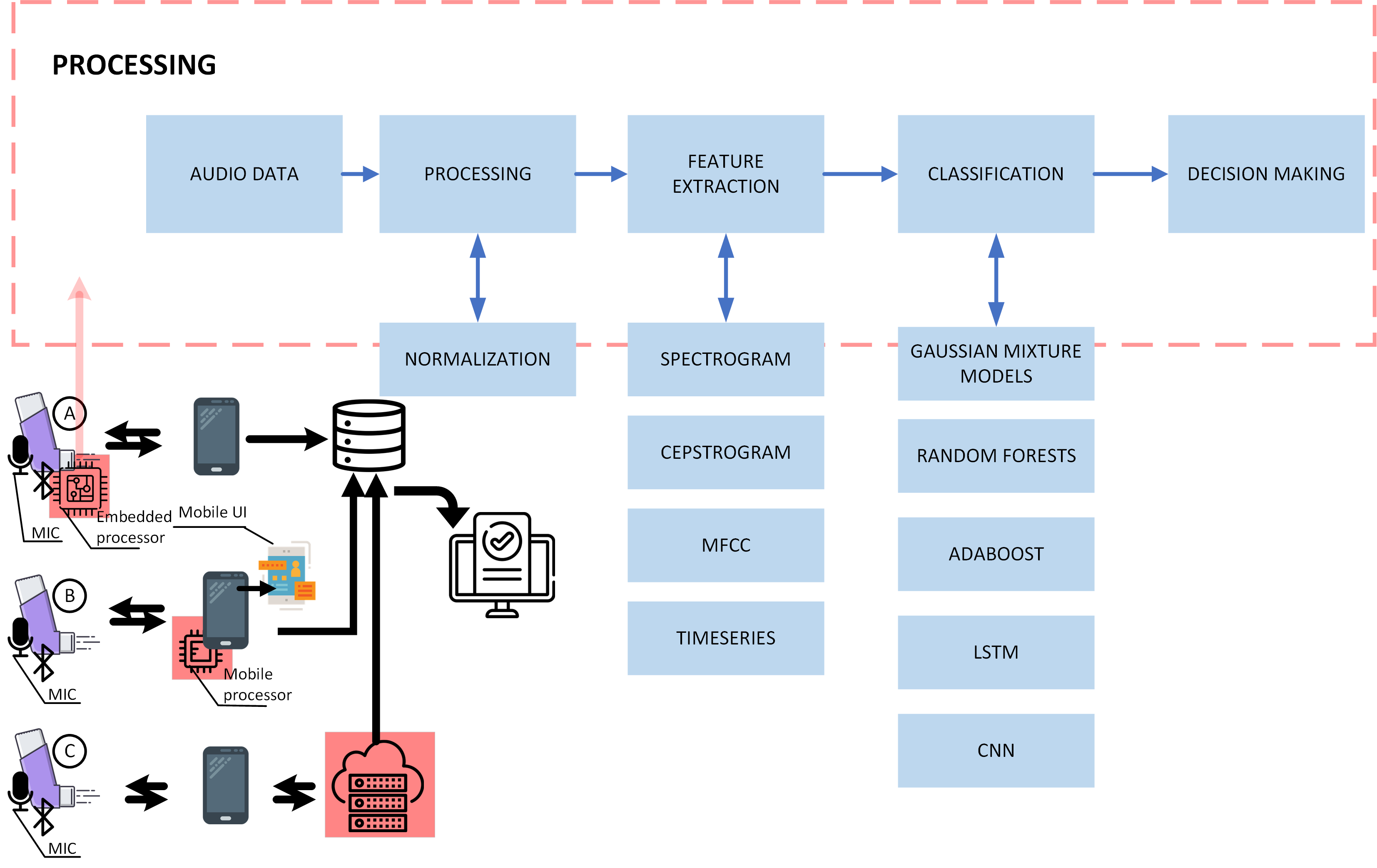}
    \caption{Flow Diagram}
    \label{fig:flowdiagram}
    \vspace{1em}
\end{figure*}

The tool for data annotation was the Audacity, whereas feature extraction and signal analysis were performed using Python Libraries (Pandas and Numpy). The overall flow of research and individual methodological components on inhaler's and respiratory sound's analysis, are illustrated in figure \ref{fig:flowdiagram}.

The rest of the paper is organized as follows: Section \ref{section:hardware} provides a synopsis on the state of the art of inhaler devices with different system designs. Section \ref{section:dataset} presents the annotation on RDA dataset recordings, after the feature extraction procedure, which is constituted in section \ref{section:featureextraction} and has been used on different approaches and methodological aspects for audio recordings. Section \ref{section:classification} analyzes the different frameworks and architectural components and the methodological aspects, which are used for the detection and recognition of respiratory and actuation sounds from RDA dataset. Section \ref{section:comparison} describes the experimental evaluation of each structure and their accuracy comparisons, while section \ref{section:future} gives recommendations for future directions on acoustic analysis and, finally, section \ref{section:conclusions} concludes this study.

\section{Inhaler technique monitoring systems}
\label{section:hardware}

There has been an increasing interest from researchers and software architects for medication adherence monitoring devices. The first inhalation device used for asthma was the pressurized metered-dose inhaler, in the 1950s. Today, there are many devices available with different techniques to support the proper drug intake \cite{masoli2004global}. We briefly present representative technologies utilized over time:
\vspace{0.5cm}
\begin{itemize}
\item SmartMist TM is a microprocessor-controlled device, widely used in academic research that optimizes drug deposition in the lung, emitted from metered-dose inhalers (MDIs) \cite{farr1995aerosol, warren1996comparison, rau2005determinants}.
\item Diskus Adherence Logger (DAL) is an inhaler device with a small size sensor, designed to identify the motion of the dose delivery level, in Diskus DPIs and to communicate with the event recorder chip, for control and data uploading to a computer \cite{bogen2004adherence, bogen2005reliability}.
\item The SmartTrack is an innovative adherence monitoring device, for pressurized metered-dose inhalers, that consists of an LCD screen and four push buttons that allow the navigation in the device menu \cite{foster2012reliability, chan2017electronic}.
\item The SmartTurbo (Adherium (NZ) Ltd, Auckland, New Zealand) is an electronic monitoring device that combines its use with a Turbuhaler device (AstraZeneca, UK) and consists of electromechanical sensors to identify the state on the mouthpiece of the inhaler \cite{gradinarsky2014inhalation, pilcher2015three, d2014method}.
\item The Asthmapolis system relies on technology that monitors the location of blister actuations, allowing the user to gain information about the disease, such as date and time of the usage \cite{van2013remote} and to collect timely and geographically specific information about asthma management, with a clear picture of health status \cite{aldridge2011project, van2013monitoring}.
\item The "Inspiromatic" is an innovative approach to inhaler enhancement, based on the real-time inhalation flow measurements \cite{shakshuki2017improving}.
\item Sensohaler is a novel device that incorporates MDIs, with fundamental acoustic sensing functionalities that are used for the prediction of volumetric flow rate \cite{denyer2010adherence}.
\item The T-Haler device is based on another innovative approach for the design of a MDI, with enhanced monitoring key performance characteristics \cite{shakshuki2017improving}.
\item Furthermore, an integrated system was presented at the University of Patras \cite{nousias2016monitoring}, consisting of three main parts: the monitoring device, the smartphone application and the cloud processing server part.
\end{itemize}
\vspace{0.5cm}

\section{Evaluation dataset}
\label{section:dataset}

\begin{figure}
    \centering
    \includegraphics[width=\linewidth]{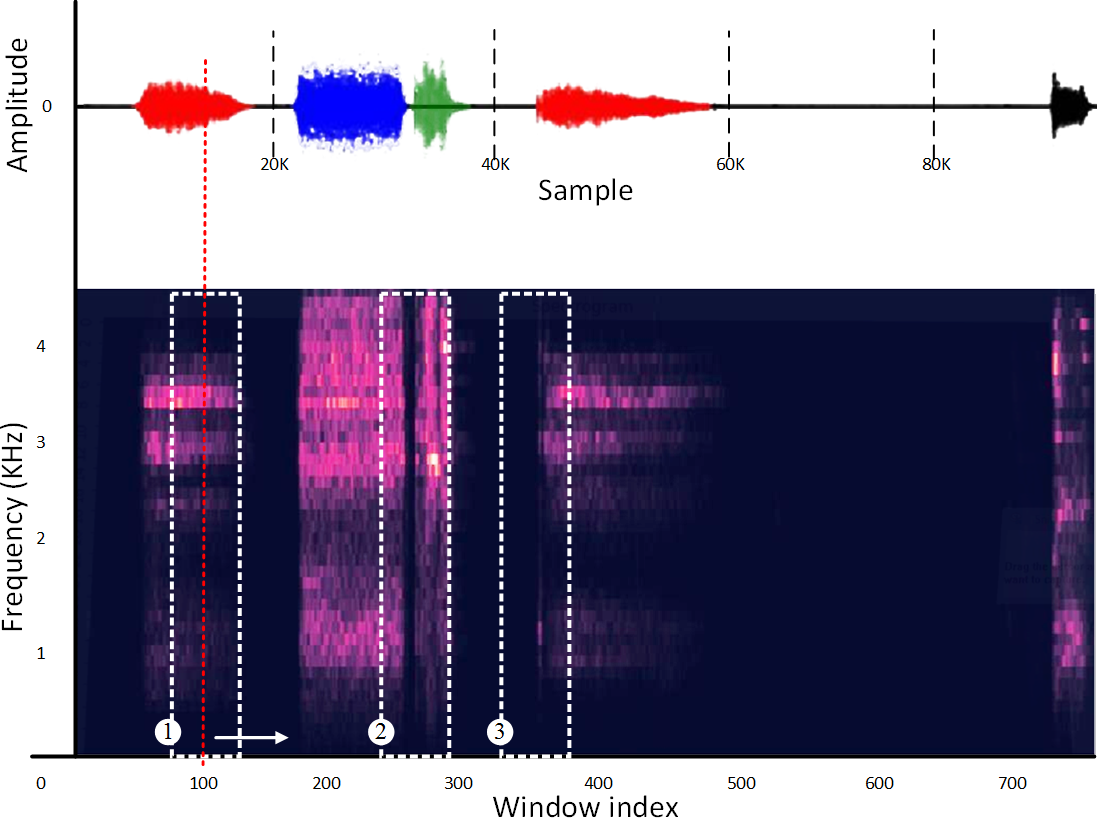}
    \caption{Audio timeseries colored with ground truth labels (top) and visualization of the corresponding spectrogram, within a sliding window (bottom). Events include inhalation (blue), drug actuation (green), exhalation (red) and environmental noise (black). Window sliding positions might include non-mixed states (window 1) or transitional states, e.g. from inhalation to drug actuation (window 2) and breath holding to exhalation (window 3), respectively. The class labels of the sliding windows correspond to the central points (reproduced with permission from Pettas et al. \cite{pettas2019recognition}).}
    \label{fig:spectrogram}
\end{figure}

The central aim of this research is to identify associations between high-level classification labels and low-level features, extracted from audio clips of different semantic activities. We investigate the clinical applicability of different audio-based signal processing methods, for assessing medication adherence. The dataset \cite{g84h-ma24-22} consists of recordings acquired in an acoustically controlled setting, free of ambient indoor environmental noise, at the University of Patras. Three subjects familiar with the inhaler technique, participated in the study. The participants were instructed to use the inhaler, as typically performed in a clinical procedure. For each and every participant informed, consent was obtained. During breathing and drug actuation, the audio signals were acquired by a microphone attached to the inhalation device, communicating with a mobile phone via Bluetooth. The addition of the adherence monitoring device did not impact the normal functioning of the inhaler, which had a full placebo canister. In total, 370 audio files were recorded with different duration each, containing an entire inhaler use case, with respiratory flow ranging on 180-240 L/min. Each audio recording was sampled with a 8KHz sampling frequency, as a mono channel WAV file at 8-bit depth. The audio recordings were segmented and annotated by a human specialist into inhaler actuation, exhalation, inhalation and environmental noise. The obtained segments (of non-mixed states) were of variable length and, for some methods, were further segmented into frames of fixed length for the purposes of feature extraction, as described in section \ref{section:featureextraction}. The acoustic signal of a typical patient recording is shown in figure \ref{fig:spectrogram}. The constructed database overall consisted of 193 drug actuation segments, 319 inhalation, 620 exhalation and 505 environmental noise segments, ready to be used for audio recognition, using different sets of features.\par

\section{Preliminaries on feature extraction and audio descriptors}
\label{section:featureextraction}

Various features have been extracted from audio signals, both in temporal and spectral domain and have been used as a basis for audio analysis algorithms \cite{duong2015review}. It is typical for audio analysis to extract the features across sliding time windows, in order to capture the class or activity within that particular moment in time. The extraction and selection of robust and descriptive features for specific applications, is the main challenge in designing audio classification systems \cite{zhang1998hierarchical, heryanto2011direct}. We have elaborated our work on the characterization of the audio signals, using several spectral features and audio patterns, as proposed in the literature. We first denote a signal $x\left(t\right), t=1,...,T$ as a time series in time domain. As the signal is changing through time, it is assumed that on short time scales, the first is statistically stationary and, thus, statistical features can be extracted through a windowing process, in which the signal is segmented into small, possibly overlapping, time windows (frames) of the same length ($N$). We denote as $\mathrm{x}_{\mathrm{n}}\left(i\right)$ the $i^{th}$ sample in the $\mathrm{n}^{th}$ frame audio signal, where $i=1,...,N$.

\subsection{Volume in time domain analysis}

\begin{figure}[h]
    \centering
    \includegraphics[width=\linewidth]{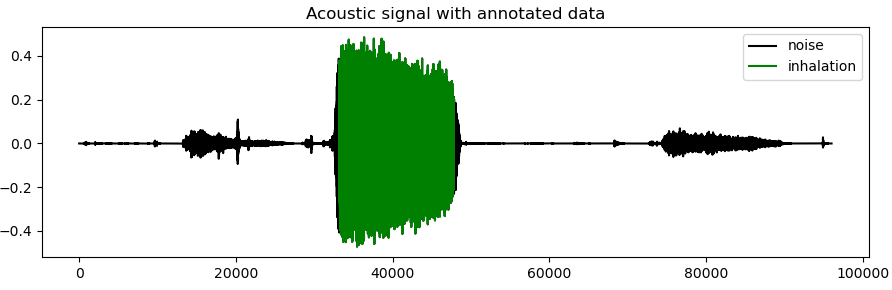}
    \caption{Annotated inhalation segment in a sound wave plot}
    \label{fig:sound_volume_inhalation}
\end{figure}

Volume in the domain of time is a reliable indicator for silence detection. Therefore, it can segment audio sequences and determine clip boundaries. It is commonly perceived as loudness, since natural sounds are pressure waves with different amounts of power that modifies the signal, with different distributions for each audio recording. In electronic sounds, the physical quantity is amplitude and, therefore, volume is often calculated as the Root-Mean-Square (RMS) of amplitude. Volume of the n$^{th}$ frame is calculated by the following formula:
\begin{equation}
\mathrm{V}(\mathrm{n})=\sqrt{1 / N} \sum_{i=1}^{N} \mathrm{x}_{\mathrm{n}}^{2}\left(i\right)
\end{equation}
\noindent where $\mathrm{V}(\mathrm{n})$ is the volume, at $n$ point in time domain. Figure \ref{fig:sound_volume_inhalation} shows the volume of an inhalation recording.

\subsection{Zero Crossing Rate analysis}

Zero-Crossing Rate ($\mathcal{ZCR}$) is defined formally as the number of time-domain zero-crossings, according to the pressure on the sound waves, within a defined region of the signal, divided by the number of samples of that region \cite{gouyon2000use}. In the context of discrete-time signals, a zero crossing is said to occur, if successive samples have different algebraic signs. The zero-crossing finds the rate at which the signal changes from positive to negative and vice-versa \cite{bachu2010voiced, bachu2008separation}. In some cases, only the "positive-going" or "negative-going" crossings are counted, rather than all the crossings, since, for a deterministic reason, between a pair of adjacent positive zero-crossings, there must be one negative zero-crossing \cite{shete2014zero}. This feature has been used extensively in speech recognition and music information retrieval, to classify percussive sounds:

\begin{align*}
\mathcal{ZCR} = \frac{1}{T-1}\sum_{t=1}^{T-1} J\{x\left(t\right)x\left(t-1\right)<0\}
\end{align*}

\noindent where $x\left(t\right)$ is the signal on time domain, with length $T$ and $J\{y\}$ is a logical function returning one, if its argument is true and zero otherwise \cite{bitopi}.

\subsection{Spectrogram analysis}

Any sound signal can be expressed in the frequency spectrum, which shows the average amplitude of various frequency components in the audio signal and the (frequency) distribution. To obtain frequency domain features, the spectrogram of an audio clip in the form of Short-Time Fourier Transform (STFT), is calculated for each audio frame. The spectrogram is used for the extraction of two features, namely frequency centroid and frequency bandwidth.

\begin{figure}[h]
    \centering
    \includegraphics[width=\linewidth]{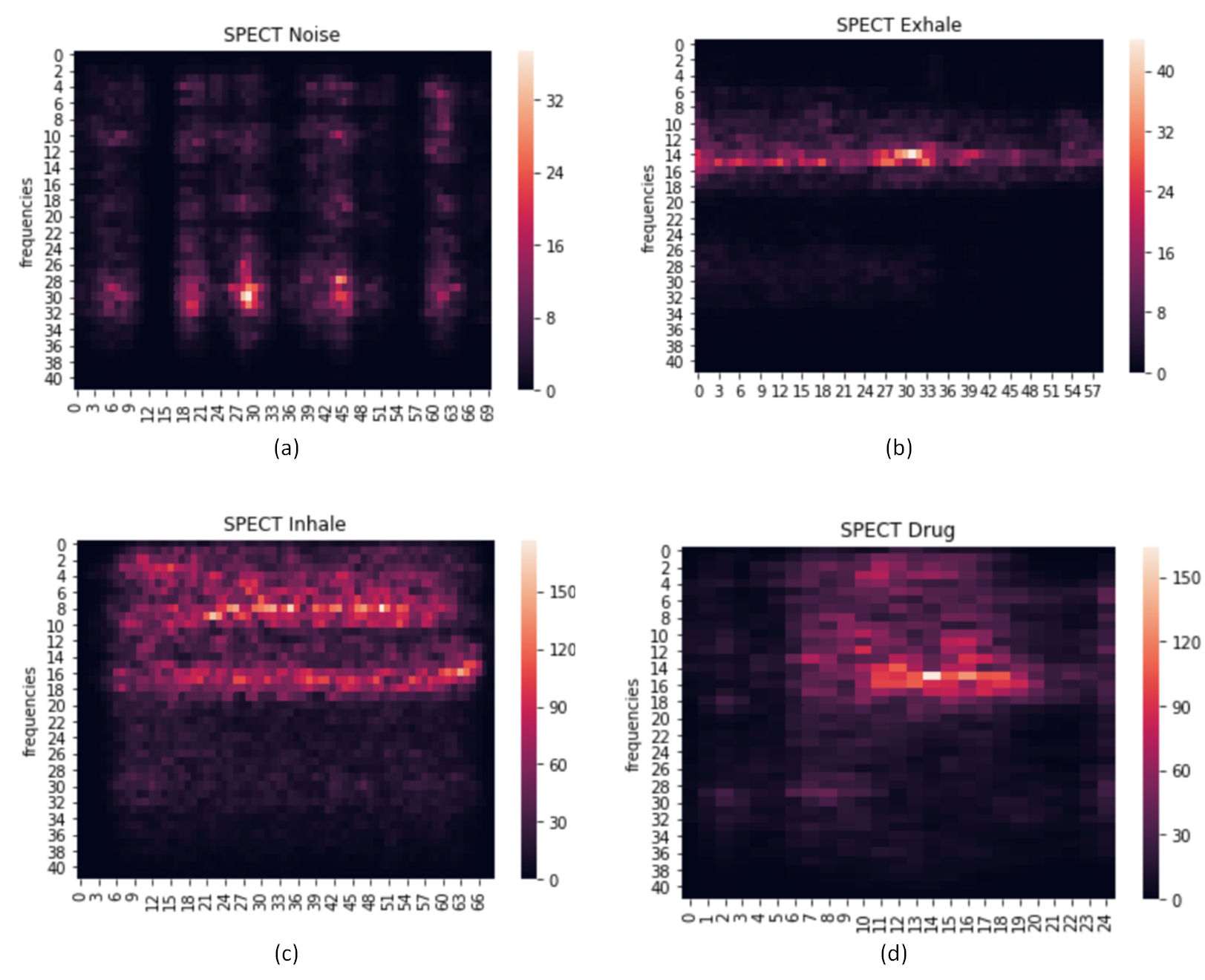}
    \caption{Spectrogram visualization for different classes (a. Environmental noise, b. Exhalation, c. Inhalation d. Drug actuation).}
    \label{fig:spectrogram_inhalation}
\end{figure}

\subsection{Cepstrum analysis}

The first paper on cepstrum analysis \cite{bogert} defined the cepstrum as "the power spectrum of the logarithm of the power spectrum". The cepstrum results from the Inverse Discrete Fourier Transform (IDFT) of a signal's log magnitude of the Discrete Fourier Transform (DFT). It has been used in speech analysis for determining voice pitch (by accurately measuring the harmonic spacing), but also for separating the formants (transfer function of the vocal tract) from voiced and unvoiced sources, which led quite early to similar applications in mechanics \cite{oppenheim}. The definition of the complex cepstrum is:

\begin{equation}
C_c\left(\tau\right) = \mathscr{F}^{-1}\left\{\log \left( F \left( f \right) \right) \right\} = \mathscr{F}^{-1}\left\{\ln \left( A\left( f \right) \right) + j\phi\left( f \right)\right\}
\end{equation}

\noindent where $\mathscr{F}^{-1}$ is the IDFT and $F$ is the DFT, in terms of the amplitude and phase of the spectrum \cite{robert}.

\subsection{Mel Frequency Cepstral Coefficients Analysis and Power Spectral Density}

Audio feature extraction is used to decrease the dimensionality of the input vector, while maintaining the discriminating power of the signal. It is an important part to prepare data for the sound identification process and this kind of analysis is derived from the cepstral representation of the audio data \cite{ganesh}. A systematic study of various spectral features can be found in (Kinnunen 2004) \cite{kinn}. The Mel Frequency Cepstral Coefficient (MFCC) feature extraction is a technique in sound recognition that is based on the frequency domain of Mel scale, for human ear scale \cite{ruinskiy}. MFCC's extraction is a significant technique, mainly, due to efficient computation schemes and its robustness in the presence of different noise \cite{sahidullah2012design}. To compute the MFCC's, a Hamming window is multiplied with the overlapping segments, after windowing and the Fast Fourier Transform (FFT) is computed for every frame. The equation for Hamming window sequence can be defined by:

\begin{equation}
w\left(n\right) = \alpha - \beta\cos(\frac{2 \pi n}{N-1})
\quad \textrm{for} \quad
-\frac{N-1}{2}\leq n \leq \frac{N-1}{2}
\end{equation}

\noindent with $\alpha=0.54$ and $\beta=1-\alpha=0.46$. For each frame we take the periodogram-based Power Spectral Density (PSD) estimation:

\begin{equation}
\hat{P}_i\left(k\right) = \frac{1}{N}\abs{X_i\left(k\right)}^2
\end{equation}

where $X_i\left(k\right)$ is the complex DFT of the signal $x\left(n\right)$, with $N$ samples:

\begin{equation}
X\left(k\right) = \sum_{n=\frac{N}{2}-1}^{\frac{N}{2}}x\left(n\right)\exp\left(-j2 \pi k\frac{n}{N}\right)
\end{equation}

The power spectral density function helps calculate the total power contained in each spectral component of a specific signal. Power spectrum of any time-domain signal $x\left(t\right)$, helps to determine the distribution of the variance of data $x\left(t\right)$, over the frequency domain, in the form of spectral components, into which the actual signal can be decomposed \cite{saini2015power}. This is motivated by the human cochlea (an organ in the ear), which vibrates at different spots depending on the frequency of the incoming sounds. According to the location in the cochlea that vibrates, different nerves fire, informing the brain that specific frequencies are present. Therefore, the estimated PSD of the signal is related to the modulus of its DFT \cite{nasser}.

\begin{figure}
    \centering
    \includegraphics[width=\linewidth]{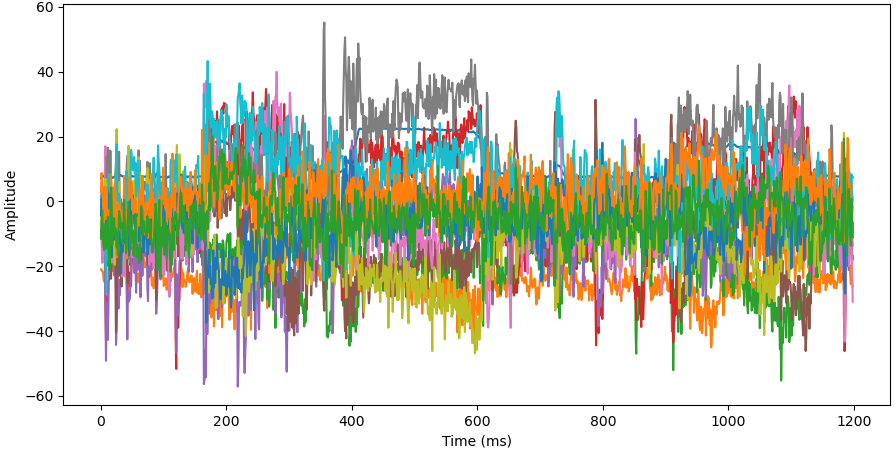}
    \caption{Mel Frequency Cepstral Coefficients of an inhalation recording.}
    \label{fig:mfcc_1_rec}
\end{figure}

After this step, we continue the calculations, with the spectrum segmented into a number of critical bands employing filterbanks. The filterbank, typically, consists of overlapping triangular filters, which are spaced linearly in a perceptual Mel scale. Then, the Mel filterbanks are calculated, in order to examine how much energy exists in various frequency regions. The Mel scale determines how to space the filterbanks and how wide to make them. It relates the perceived frequency of a pure tone to its actual measured frequency. Once we have the filterbank energies, we compute their logarithm. The logarithm allows us to use cepstral mean subtraction, which is a channel normalization technique. Finally, Discrete Cosine Transformation (DCT) is applied to the logarithm of the filterbank outputs, which results in the raw MFCC vector \cite{sirko}. Because our filterbanks are all overlapping, the filterbank energies are quite correlated with each other. The DCT decorrelates the energies, so as diagonal covariance matrices can be used to model the features in various classifiers (figure \ref{fig:mfcc_1_rec}).

There can be variations on this process, for example, differences in the shape or spacing of the windows used to map the scale \cite{zheng2001comparison} or addition of dynamic features, such as "delta" and "delta-delta" (first- and second- order frame-to-frame difference) coefficients. Exponentiating the log Mel-filter bank spectrum before the cepstrum computation, can significantly reduce the sensitivity of the cepstra to spurious low energy perturbations \cite{tyagi2005desensitizing}. The difference between the cepstrum and the Mel frequency cepstrum is that in the MFCC, the frequency bands are equally spaced on the Mel scale, which approximates the human auditory system's response more closely, than the linearly-spaced frequency bands, used in the normal cepstrum. This frequency warping can permit for better representation of sound, in audio compression. MFCC's have been successfully used in speech and music applications, playing a central part in recent efforts, to complete machine audition \cite{chad}.

\subsection{Wavelet Transform analysis}

The wavelet transform can construct a time-frequency representation of a signal, that offers very good time and frequency localization. The wavelet transform, with the wavelet $\psi$ of a signal $x(t)$ is defined as:

\begin{equation}
W_{\psi}^{x}(a, b)=\frac{1}{\sqrt{c_{\psi}|a|}} \int_{-\infty}^{\infty} x(t) \psi\left(\frac{t-b}{a}\right) d t
\end{equation}

\noindent where $a$ is called dilation and $b$ translation parameter. The Morlet wavelet, consisting of a sinusoid multiplied by a Gaussian window, is commonly used, because its scale-frequency relationship requires less computation, as the peak frequency is equal to the center frequency of the wavelet. The Morlet wavelet $\psi$, which is sometimes called Gabor wavelet, has impressive mathematical and biological properties and is given by:

\begin{equation}
\psi(t)=\frac{1}{\pi^{\frac{1}{4}}}\left(e^{i 2 \pi f_{0} t}-e^{-\left(i 2 \pi f_{0}\right)^{2} / 2}\right) e^{-t^{2} / 2}
\end{equation}

\noindent where $f_{0}$ is the center frequency of the mother wavelet \cite{bussow2007algorithm}.

\subsubsection{Continuous Wavelet Transform}
 
In definition, the Continuous Wavelet Transform (CWT) is a convolution of the input data sequence, with a set of functions generated by the mother wavelet. It provides an over-complete representation of a signal by letting the translation and scale parameter of the wavelets, vary continuously \cite{reddy2014object, selesnick2005dual}. The convolution can be calculated in the time-domain or in the frequency-domain, by using an FFT algorithm.

For the STFT, a fixed width segment size controls the time-frequency resolution trade-off. This results in a single resolution in the domain of time and a single resolution in the frequency domain, regardless of the rendered frequency. In contrast, wavelet analysis is a multi-resolution method. The time-frequency resolution is not constant, but varies with frequency. Multi-resolution analysis was designed for the common condition, where high-frequency components exist for a short time duration within a signal, while low-frequency components are more persistent. Short-lived high frequency components need strong time localization. In order to achieve this, the frequency resolution of high frequency components will be diminished. On the other hand, long-lived low frequency components can tolerate poorer time resolution, but require effective frequency resolution. Low-frequency components, often determine the significant part of a signal's character and these properties will be best quantified, if the frequency resolution is as satisfactory as possible. More references on the CWT can be found in appendix \ref{appendix:CWT}.

\section{Respiratory and Inhaler's Sound Classification}
\label{section:classification}

The elements of the pipeline will be executed in parallel and in a time-sliced fashion, at some points. In some approaches, the data must be transformed to take advantage of the value it can deliver, so it can continuously improve the accuracy of the models and achieve successful results \cite{mcfarland2006bci}. We tried to have the appropriate data quality, reliability and accessibility. Metadata extraction is correlated with the captured data and provides descriptive and targeted information, about the object and the data itself.

\subsection{Classical Machine Learning Algorithms for Feature-based Respiratory Signals Classification}

\subsubsection{Decision Trees with CWT}

Chronologically, the first algorithm that was employed to identify actuation sounds \cite{taylor2014acoustic} was utilized Decision Trees (DTs) with CWT, for inhalation classification. The CWT was calculated using a Morlet Wavelet (MW) with an adjustable parameter of $20$. A peak assessment routine was then employed to detect and assess peaks, exceeding a threshold $\theta_1=0.38$. Each peak was initially marked as a potential actuation plume. Power values were taken at specific time distance ($56$ ms), before and after the peak point, to observe if the power decreased by a given threshold ($\theta_2=0.25$), in comparison to the peak value, to flag it as an actuation plume. It was observed that the actuation acoustic signal was of very small time duration (100-150ms). The wavelet variance is calculated, so that different datasets may be compared at different scales:

\begin{equation}
V(a)=\frac{1}{n} \sum_{j=1}^{n} W^{2}\left(a, x_{j}\right)
\end{equation}

\noindent where $W^{2}\left(a, x_{j}\right)$ is the squared wavelet coefficient, associated with scale $a$ at data point $x_{j}$ and $n$ is the number of data points. From this definition, wavelet variance is a function of scale. Bradshaw and Spies \cite{bradshaw1992characterizing} pointed out that "high values of wavelet variance, at a given scale, reflect the presence of a greater number of peaks or a greater intensity of the signal, or both".\par

Other algorithmic approaches in the sub-field of blister and respiratory sound detection and classification, follow a 3 stages methodology including (i) blister detection, (ii) breath detection and (iii) inhalation/exhalation differentiation \cite{holmes2014acoustic, taylor2018advances}. The algorithms were designed and developed to automatically detect inhaler events from the audio signals and provide feedback, concerning medication adherence. These approaches have multiple clinical implications, as they prove the practicability of using acoustics to objectively monitor patient inhaler adherence and provide real-time personalized medical care for chronic respiratory illness.\par

\subsubsection{Hidden Markov Model}

Another approach proposes the Directed Acyclic Graph (DAG) logic \cite{ntalampiras2019automatic}. The formulation relies on a graph denoted as $\mathcal{G}=\{N, L\}$, where $N=\left\{n_1, \ldots, n_m\right\}$ represents the nodes and $L=\left\{l_1, \ldots, l_p\right\}$ the links. Each node in $N$ is responsible for a binary classification task conducted via a set of Hidden Markov Models (HMMs), which fit well the specifications for sound pattern classification \cite{ntalampiras2021acoustic}. DAGs can be seen as a generalization of the class of Decision Trees (DTs), while repetitions that may occur in different branches of the tree can be handled more efficiently, since different decision paths might be merged. In order to get an early indication of the degree of difficulty of a classification task, a metric is employed representing the distance of the involved classes in the probabilistic space, the Kullback-Leibler Divergence (KLD). The KLD between two J-dimensional probability distributions A and B is defined as:

\begin{equation}
D(A \| B)=\int_{R^J} p(X \mid A) \log \frac{p(X \mid A)}{p(X \mid B)} d x
\end{equation}

\subsubsection{Quadratic Discriminant analysis}

In 2018, a Quadratic Discriminant Analysis (QDA) model was employed, for audio-based analysis of respiratory sounds \cite{taylor2018objective}. The algorithm was composed of two phases, training and testing, to automatically recognize the sound events. In both training and testing phases, the inhaler audio signal was band-pass filtered at $140$ to $22000$ Hz, to emphasize the events and to reduce the external noise. The audio signals were divided into frames of $40$ ms duration, with $20$ ms overlap. The DC offset (mean amplitude displacement from zero) was removed from each frame. Thirty audio-based features from time and spectral domains were extracted, for each frame. The extracted features are: $12$ MFCC's, $10$ Linear Predictive Coding (LPC) coefficients, PSD, ZCR and a high frequency power value (over $15$ kHz), estimated using the CWT. Since the Flo-Tone device generates a harmonic sound during inhalation, a harmonic feature was also extracted. This harmonic feature was calculated as the peak value of the frame's auto-correlation function, searched in the range of 500–600Hz.

Classifying between two multivariate normal populations leads to the idea of discriminant analysis, which is essentially the Bayes classifier for the problem, that constructs a combination of the features. The features from two classes follow multivariate normal distributions, with different means $\mu_{i}$, with precision matrices $\Omega_{i}=\Sigma_{i}^{-1}$, with $i = {1, 2, ..., Q(Y)}$, where $Q(Y)$ is the cardinal number of the set $Y$ of the classes. In this, high dimensional setting, these matrices can be estimated only in presence of sparsity.

Consider observing training data $\left(X_{m}, Y_{m}\right), m=1, \ldots, n$, where $X_{m} \in \mathfrak{X}$ and $Y_{m} \in\{0,1, .., n\}$ and
$$
X_{m} \mid\left\{Y_{m}=k\right\} \sim p_{i}(x), \text { if } Y_{m}=i,
$$
for density $p_{i}$. A classification rule (or a classifier) $\phi$ is a map $\phi: \mathfrak{X} \rightarrow[0, ..., Q(Y)]$, such that $\phi(x)$ stands for the probability of classifying a new observation to Class $i$. The misclassification rate of a classifier $\phi$ is given by $r(\phi)=\mathrm{E}[Y(1-\phi(X))+(1-Y) \phi(X)]$.

The idea of the one-versus-the-rest method is as follows: to get a $K$-class classifier, first construct a set of binary classifiers $C_{1}, C_{2}, \cdots, C_{K}$. Each binary classifier is first trained to separate one class from the rest and, then, the multi-class classification is carried out according to the maximal output, of the binary classifiers. Since the binary classifiers are obtained by training on different binary classification problems, it is unclear whether their real-valued outputs (before thresholding) are on comparable scales \cite{scholkopf2002learning}. In practice, however, situations often arise, where several binary classifiers assign the same instance to their respective class or where none does \cite{scholkopf2002learning}.

\subsubsection{Support Vector Machines}

Support Vector Machine (SVM) classifier is based on statistical learning theory \cite{vapnik1998support} and has shown to be one of the most robust supervised learning methods. SVMs' simplicity comes from the fact that they apply a simple linear method to the data, but in a high-dimensional feature space are non-linearly related to the input space. In binary classification, only the decision boundaries of the first class are to be known and the rest (complement of first-class) is considered as the second class, whereas in multi-class classification, several decision boundaries need to be calculated, which may lead to increase of error probability. SVMs are highly accurate and able to model complex non-linear decision boundaries. This classifier may be applied both to linearly and non-linearly separable data, with the use of kernel transformations. Specifically, it transforms the data to a higher dimension, from where it can identify a hyperplane that separates the data.

Furthermore, an other interesting application of SVM for respiratory signal classification is presented in Eleftheriadou et al. \cite{eleftheriadou2020audio}, where an audio-based method is assessing the proper usage of dry powder inhalers, by using the FFT of the signal. A window size of $512$ frames, with a number of frames between FFT columns ($128$ frames), was used and $16$ MFCC's including the zero coefficient, were returned resulting in an array of size ($16 \times 126$). For "silent" areas removal, the short-term features of the whole recording were extracted and an SVM model was trained to distinguish between high-energy and low-energy short-term frames, using 50\% of the highest and 50\% of the lowest energy components, for SVM model training. Then, the segments containing sound and the active segments were detected by median and dynamic thresholding, respectively. In addition to this model, a work in 2021 showed that the SVM classifier can be used on inhalation sounds and achieved an accuracy of $96.9$\%, using an open access respiratory sounds database \cite{rocha2020automatic}.

\subsubsection{Random Forest}

The Random Forest (RF) algorithm trains several tree-like classifiers \cite{moham} and aggregates results, by majority voting. The RF usually illustrates high accuracy and processing speed, though the correlation or/and independence of trees may affect the accuracy of the outcomes. The RF classifier draws $n$ tree bootstrap samples from the original data and, among the variables, the best split is selected. The accuracy of the RF overall depends on the strength of each tree and the correlation between any two trees. Each tree in RF can also be constructed by a bootstrap sample from the data, using a small set of randomly selected attributes.

\subsubsection{Adaboost}

Boosting was proposed by Freund and Schapire, in 1990 \cite{freund1999short}. Adaboost is the most common boosting algorithm. It is an efficient instrument for improving the predictive ability of a learning system and a most typical method in coordinating learning. It, usually, employs DT models as weak learners and evaluates them sequentially \cite{hastie2009multi}. Subsequent DTs are updated in favour of those samples misclassified by previous DTs.

Preparing training sets $\left(x_{1}, y_{1}\right), \ldots,\left(x_{n}, y_{n}\right)$, where $x_{i} \in X$ and $X$ represents a certain domain or instance space and each member is a training example, with a label. In the initial development, it was a two-class problem of learning. In going from two-class to multi-class classification, most boosting algorithms have been restricted, reducing the multi-class classification problem to multiple two-class problems (e.g. as shown in early works \cite{freund1997decision, schapire1997using, schapire1999improved, allwein2000reducing, friedman2000additive, friedman2001greedy}). The weights of all the training examples are initially set to be $1 / \mathrm{m}$ equally. Adaboost conducts $T$ times of iteration through repeatedly calling a weak learning algorithm. As per $D_{t}$ distribution, the weak learner finds appropriate weak hypothesis $h_{t}: \quad X \rightarrow R$, thus, predicting function sequence is gained, at this point. In the simplest case, if the scope of each $h_{t}$ is two-valued $\{-1,+1\}$, the task of the learner is to minimize the error. Combining $T$ weak hypotheses, the final predicting function (hypothesis) $H$ is gained, after $T$ times of circulation, with a weighted majority voting method. The learning accuracy rate of a single weak learner is not high enough \cite{chengsheng2017adaboost}, however, with the application of Boosting algorithm, the accuracy rate of the final result is to be improved. The Adaboost algorithm is described in detail, in appendix \ref{appendix:Adaboost}.

\subsubsection{Gradient Boosting}

Recent work on inhalation signal classification \cite{eleftheriadou2020audio} has used Gradient Boosting, as a numerical optimization method. The objective has been to minimize the loss function by adding weak learners in a gradient descent type procedure. Decision trees have also been used as weak learners in gradient boosting to produce an ensemble of weak prediction models \cite{schapire2003boosting}. The DT model is built in a stage-wise fashion and the generalization is achieved by an arbitrary differentiable loss function. The Gradient Boosting algorithm can easily overfit a training data set and that is why different regularization methods are applied to improve the algorithm's performance and address the problem of possible overfitting \cite{bartlett2006adaboost}.

\subsection{Gaussian Mixture Models}

Gaussian Mixture Models (GMMs) can be employed to approximate any Probability Density Function (PDF), given a number of components. It is proposed a novel content based audio classification approach, for monitoring pMDI medication adherence \cite{nousias2018mhealth}, that exploits the separability of the cepstrogram features, using a GMM classifier. GMMs are statistical models for representing normally distributed sub-populations, within an overall population and are used in many pattern recognition applications. For each class, a separate model is trained by fitting the corresponding feature vectors to a GMM, with parameters $\{a_i, \mu_i, C_i\}$, where $i \in \{1,...,K\}$ and $K$ is the number of components, $a_i$ is the mixture weight of component $i$, $\mu_i$ is the d-dimensional vector containing the mean values for each feature and $C_i$ is the corresponding co-variance matrix. The Gaussian mixture density $p\left( \nu | \lambda_n \right)$ is modeled as a linear combination of multivariate Gaussian PDFs, where $\nu$ is a feature vector and $\lambda_n$ is the GMM, corresponding to class $n$. In order to classify a test feature vector, we derive the $P\left(\nu|\lambda_n\right)$ for each class. The test feature vector is assigned to the class $n$, with the greatest likelihood $P\left(\nu|\lambda_n\right)$. An expectation maximization (EM) approach is utilized to derive the parameters $K_n$, $\{a_i, \mu_i, C_i\}$ for the GMM $\lambda_n$, corresponding to the class $n$ that best fits the input data.\par

After the optimal parameters for the GMMs have been computed and given $d$ the number of features, $K$ the number of components of the $i^{th}$ feature vector $v_i$ and $\lambda_n$ the GMM of class $n$, we have the following equation:

\begin{equation}
    P\left(v_i|\lambda_n\right) = \sum_{i=1}^{K}{{a_i}^n{p_i}^n{\left(v\right)}}
\end{equation}

where ${a_i}^n$ are the mixture weights to satisfy the constraint:

\begin{equation}
    \sum_{i=1}^{M}{{a_i}^n} = 1, \quad {{a_i}^n} \leq 0
\end{equation}

Finally, after the $P\left(v|\lambda_n\right)$ for the test feature vector $v$ and for each class $n$ is estimated, the test feature vector is assigned to the class $n$ with the greatest likelihood. Then, we have the relevant feedback. This procedure lies in the assumption that the initial dataset was compiled by a small group of people. For the derivation of the GMM, through the EM approach, we refer to the appendix \ref{appendix:GMM}.

\subsection{Deep Learning Models}

Neural Networks (NNs) have been employed in the past for a variety of classification problems and have shown significantly accurate results, in medical applications \cite{geman2013nonlinear, kiranyaz2015real} and general audio classification problems \cite{lee2009unsupervised, abdel2012applying}. The Convolutional Neural Networks (CNNs) can adapt to the characteristics of the training dataset and create a hierarchy of increasingly complex features \cite{ji20123d}, while at the same time, they illustrate relatively fast and consistent convergence in the training process. This model is robust in automatically learning the intrinsic patterns from the data, which can both prevent time-consuming manual feature engineering and capture hidden intrinsic patterns more effectively. Moreover, a CNN is more capable of discovering intricate patterns in high-dimensional data, compared with manual feature engineering. A pooling layer is introduced in each stage to merge similar features, reducing the dimensionality and dealing with some motif variations due to small signal variability (shifts and distortions).

The prior probabilities of each class were determined by their frequency in the training dataset. A function that takes audio recordings as inputs was developed, which returned segment endpoints that correspond to individual sound events. Backpropagation was used to train the multi-layer neural network models, which computes the gradient of a predefined objective function with respect to all the neuron parameters. The gradients are propagated backwards from the output layer to the input layer to adjust the parameters, such that the network can converge to a state that is able to encode the training patterns.

\subsubsection{Convolutional Neural Network's approaches}

In classical machine learning based direction, it is proposed an algorithmic methodology, using CNNs, intending to produce more accurate results, for real life environments, as opposed to controlled laboratory conditions with reduced levels of noise. This type of models has been established as a reliable, data-driven approach for time series and image classification \cite{hussain2018swishnet, kopuklu2019real}. Also, the authors in \cite{krizhevsky2012imagenet} have demonstrated their accuracy and efficiency in classification tasks, while strengthening the basis for CNN architectures.

One important CNN approach \cite{kikidis2015utilizing} has a deep architecture that is consisted of two convolutional layers, of 16 1D kernels of size $100$. Two fully connected layers of size $256$ and $16$, each of which was subjected to 50\% dropout in the training process, lead to the output of two distinct states. For all layers of the CNN, RELU was selected as the activation function, in order to reduce the computational complexity of the algorithm. Furthermore, the initialization of the network's parameters was performed, using a random generator of uniform distribution for the layers. Finally the training of the 1D CNN utilized Adam optimizer and was based on the categorical cross-entropy, between the predictions and the target values.

Later, a fast data-driven approach \cite{ntalianis2019assessment, ntalianis2020} was proposed, based on 2D CNNs. The benefits according to the authors' belief, can be summarized in the following points:
\begin{itemize}
\item The presented approach is applied directly on the time domain, to gain from reduced computational complexity.
\item Convolutional deep sparse coding speeds up the computational graph, aiming to allow the real-time implementation. Specifically, the architecture utilizes three convolutional layers, with each layer constituted by a max pooling layer and a dropout function \cite{wu2015max}. The output is fed to a set of four dense layers. To compute the loss of each model, it is used categorical cross entropy. Furthermore, the stride is set to one and it was used zero padding to keep the shape of the output of each filter, constant. By processing an audio recording, a vector of $n$ samples is created, reshaped in a two-dimensional array.
\end{itemize}

Next, an ensemble of C-DNN and Autoencoder networks is deployed for the analysis and classification of respiratory sounds \cite{ngo2021deep}. In this work, the experiments were conducted using the 2017 ICBHI (Internal Conference on Biomedical Health Informatics) benchmark dataset. Also, an attention-based CNN was developed for automatic identification of respiratory illness and applied on the same database \cite{yang2020adventitious}. The experimental results indicate that the residual networks can lead to important improvement, as compared with the baseline algorithms.

\subsubsection{LSTM analysis and implementation}

In this approach \cite{pettas2019}, deep layers were added to the network. The proposed sequential architecture consists, initially, of one layer of LSTM memory cells, with each one consisting of $h=64$ units. After the LSTM input layer, a dropout layer \cite{srivastava2014dropout} is introduced, in order to reduce overfitted parts after training, with dropout rate set to $0.3$, followed by a flatten layer and a dense output layer, that returns a $4 \times 1$ vector. Finally, a softmax activation function is used. The model is optimized using binary cross-entropy loss \cite{liu2017learning} and the Adam optimizer \cite{kingma2014adam}.

The spectrogram was used as a tool to develop a classifier of inhaler sounds. It is swept across the temporal dimension, with a sliding window with length $w=15$, moving at a step size being equal to $\mathbf{1}$ window. In order to form the training instances, we segment $\mathbf{S}$ into time windows of size $N_f \times T$ and assign a class to each one of them, according to the class of the central point of the window, as presented in figure \ref{fig:spectrogram}. Each training example $\mathbf{W}_k$ is defined as:

\begin{equation}
  \mathbf{W}_k \in \mathcal{R}^{ N_F \times T}= (\mathbf{S}_{i j})_{\substack{1 \le i \le N_F\\ k-w \le j \le k}}
\end{equation}

The training set is organized in microbatches $\mathbf{B}$, so that $\mathbf{B} \in \mathcal{R}^{b\times (T\times N_F)}$ (with $b=25$ in our experiments). The input tensor can be defined as $\mathbf{X} \in \mathcal{R}^{n \times w \times F}$, where $n=25$ is the minibatch size, $w=25$ is the window size and $F=42$ is the dimension of the spectrogram feature vector.\par

Extensive hyper-parameter optimization took place, to define the number of hidden units, the number of dropout rate and the minibatch size. By observing the performance of the network on the validation set, we stopped training at 70 epochs, to avoid over-fitting. The testing loss increases and the average testing accuracy stabilizes after around 70 epochs.\par

\section{Comparative Analysis of the algorithmic approaches}
\label{section:comparison}

\subsection{Simulation setup and validation settings}

This work aims to highlight the RDA suite of methods and models \footnote{\url{https://codeocean.com/capsule/8383844/tree}} trained with the Respiratory and Drug Actuation Dataset \footnote{\url{https://ieee-dataport.org/documents/respiratory-and-drug-actuation-dataset}}.
The methodology of the simulation studies, presented in \cite{nousias2022patient}, entails the comparison of the classification performance of the aforementioned important and widely used machine learning and deep learning algorithms, namely the Random Forests (RFs), Support Vector Machines (SVMs), Adaboost, Convolutional Neural Networks (CNNs) and Recurrent Neural Networks (RNNs). 

They were evaluated using spectrogram, cepstrogram and MFCC features. CNNs were directly applied on timeseries values aiming to demonstrate their capability to provide dependable solutions at lower execution times.\par

A main differentiation parameter in validation comes from the availability of previous recordings, from a specific individual. It is expected that prior information can increase the classification accuracy, however it puts additional burden in the usage of the monitoring system, since it requires the collection of data every time a new patient wants to test the framework.\par

Firstly, we consider the \textit{Multi Subject} modeling approach, denoted as \textit{MultiSubj}. In this case, the recordings of all subjects are used to form a large dataset, which is divided in five equal parts used to perform five-fold cross-validation, thereby allowing different samples from the same subject to be used in training and test set, respectively. This validation scheme was followed in previous work \cite{ntalianis2019assessment} and thus is performed, also, here for comparison purposes.\par

The second case includes the \textit{Single Subject} setting, in which the performance of the classifier is validated through training and testing, within each subject’s recordings. We denote such models, as \textit{SingleSubj}. The recordings of each subject are split in five equal parts, to perform cross-validation. The accuracy is assessed for each subject separately and, then, the overall performance of the classifier is calculated by averaging the three individual results.

The third evaluation setting refers to the case, when no previous recordings for the testing subject are available, thus samples from other subjects are used. This is the \textit{leave-one-subject-out (LOSO)} approach that illustrates how well the trained network can generalize to individuals that it never saw before, during training. \textit{LOSO} models facilitate the use of the monitoring system, since they don't require a data pre-collection phase and, also, they have the lowest risk of over-fitting. However, if the inter-subject variability is high, the data might not adapt well, especially if the number of the training subjects is small, as in our case. With this approach we use the recordings of two subjects for training and the recordings of the third subject for testing. This procedure is completed when all subjects have been used for testing and the accuracy is averaged to obtain the overall performance of the classifiers.

Furthermore, a new parameter was brought into the comparison, that is the testing dataset mixing. A non mixed setup means that each audio segment in the testing set, consists of a single class, but in a mixed setup each testing set sample emanates from a sliding window, that naturally includes parts belonging to multiple classes. In this case, the class of an audio segment is the class of the central sample.

\begin{table}[]
\centering
\caption{Accuracy}
\label{table:results:Accuracy}
\resizebox{\linewidth}{!}{
\begin{tabular}{|l|l|l|r|r|r|}
\hline
{\color[HTML]{000000} }                             & {\color[HTML]{000000} }     & {\color[HTML]{000000} }      & {\color[HTML]{000000} Multi}                         & {\color[HTML]{000000} LOSO}                          & {\color[HTML]{000000} Single}                        \\ \hline
{\color[HTML]{000000} }                             & {\color[HTML]{000000} ada}  & {\color[HTML]{000000} cepst} & \cellcolor[HTML]{CFCFCF}{\color[HTML]{000000} 92.11} & \cellcolor[HTML]{CFCFCF}{\color[HTML]{000000} 90.35} & {\color[HTML]{000000} 90.67}                         \\ \cline{2-6} 
{\color[HTML]{000000} }                             & {\color[HTML]{000000} ada}  & {\color[HTML]{000000} mfcc}  & {\color[HTML]{000000} 91.76}                         & {\color[HTML]{000000} 88.31}                         & {\color[HTML]{000000} 89.43}                         \\ \cline{2-6} 
{\color[HTML]{000000} }                             & {\color[HTML]{000000} ada}  & {\color[HTML]{000000} spect} & {\color[HTML]{000000} 89.8}                          & {\color[HTML]{000000} 84.55}                         & {\color[HTML]{000000} 88.79}                         \\ \cline{2-6} 
{\color[HTML]{000000} }                             & {\color[HTML]{000000} cnn}  & {\color[HTML]{000000} time}  & {\color[HTML]{000000} 91.71}                         & \cellcolor[HTML]{CFCFCF}{\color[HTML]{000000} 92.94} & {\color[HTML]{000000} 90.61}                         \\ \cline{2-6} 
{\color[HTML]{000000} }                             & {\color[HTML]{000000} gmm}  & {\color[HTML]{000000} cepst} & {\color[HTML]{000000} 89.81}                         & {\color[HTML]{000000} 80.54}                         & {\color[HTML]{000000} 89.33}                         \\ \cline{2-6} 
{\color[HTML]{000000} }                             & {\color[HTML]{000000} gmm}  & {\color[HTML]{000000} mfcc}  & {\color[HTML]{000000} 82.83}                         & {\color[HTML]{000000} 82.12}                         & {\color[HTML]{000000} 78.49}                         \\ \cline{2-6} 
{\color[HTML]{000000} }                             & {\color[HTML]{000000} gmm}  & {\color[HTML]{000000} spect} & {\color[HTML]{000000} 86.45}                         & {\color[HTML]{000000} 83.58}                         & {\color[HTML]{000000} 86.88}                         \\ \cline{2-6} 
{\color[HTML]{000000} }                             & {\color[HTML]{000000} lstm} & {\color[HTML]{000000} spect} & \cellcolor[HTML]{CFCFCF}{\color[HTML]{000000} 92.16} & {\color[HTML]{000000} 87.73}                         & \cellcolor[HTML]{CFCFCF}{\color[HTML]{000000} 91.1}  \\ \cline{2-6} 
{\color[HTML]{000000} }                             & {\color[HTML]{000000} rf}   & {\color[HTML]{000000} cepst} & {\color[HTML]{000000} 91.69}                         & \cellcolor[HTML]{CFCFCF}{\color[HTML]{000000} 90.28} & \cellcolor[HTML]{CFCFCF}{\color[HTML]{000000} 91.05} \\ \cline{2-6} 
{\color[HTML]{000000} }                             & {\color[HTML]{000000} rf}   & {\color[HTML]{000000} mfcc}  & {\color[HTML]{000000} 91.61}                         & {\color[HTML]{000000} 89.36}                         & {\color[HTML]{000000} 90.86}                         \\ \cline{2-6} 
{\color[HTML]{000000} }                             & {\color[HTML]{000000} rf}   & {\color[HTML]{000000} spect} & {\color[HTML]{000000} 89.47}                         & {\color[HTML]{000000} 85.7}                          & {\color[HTML]{000000} 89.85}                         \\ \cline{2-6} 
{\color[HTML]{000000} }                             & {\color[HTML]{000000} svm}  & {\color[HTML]{000000} cepst} & {\color[HTML]{000000} 91.67}                         & {\color[HTML]{000000} 82.34}                         & {\color[HTML]{000000} 90.89}                         \\ \cline{2-6} 
{\color[HTML]{000000} }                             & {\color[HTML]{000000} svm}  & {\color[HTML]{000000} mfcc}  & \cellcolor[HTML]{CFCFCF}{\color[HTML]{000000} 92.49} & {\color[HTML]{000000} 87.29}                         & \cellcolor[HTML]{CFCFCF}{\color[HTML]{000000} 92.14} \\ \cline{2-6} 
\multirow{-14}{*}{{\color[HTML]{000000} Mixed}}     & {\color[HTML]{000000} svm}  & {\color[HTML]{000000} spect} & {\color[HTML]{000000} 34.03}                         & {\color[HTML]{000000} 34}                            & {\color[HTML]{000000} 33.53}                         \\ \hline
{\color[HTML]{000000} }                             & {\color[HTML]{000000} ada}  & {\color[HTML]{000000} cepst} & {\color[HTML]{000000} 95.02}                         & \cellcolor[HTML]{CFCFCF}{\color[HTML]{000000} 93.32} & {\color[HTML]{000000} 92.57}                         \\ \cline{2-6} 
{\color[HTML]{000000} }                             & {\color[HTML]{000000} ada}  & {\color[HTML]{000000} mfcc}  & \cellcolor[HTML]{CFCFCF}{\color[HTML]{000000} 95.91} & {\color[HTML]{000000} 87.02}                         & {\color[HTML]{000000} 93.8}                          \\ \cline{2-6} 
{\color[HTML]{000000} }                             & {\color[HTML]{000000} ada}  & {\color[HTML]{000000} spect} & {\color[HTML]{000000} 94.01}                         & {\color[HTML]{000000} 84.92}                         & {\color[HTML]{000000} 92.99}                         \\ \cline{2-6} 
{\color[HTML]{000000} }                             & {\color[HTML]{000000} cnn}  & {\color[HTML]{000000} time}  & \cellcolor[HTML]{CFCFCF}{\color[HTML]{000000} 95.29} & \cellcolor[HTML]{CFCFCF}{\color[HTML]{000000} 94.12} & {\color[HTML]{000000} 92.69}                         \\ \cline{2-6} 
{\color[HTML]{000000} }                             & {\color[HTML]{000000} gmm}  & {\color[HTML]{000000} cepst} & {\color[HTML]{000000} 94.1}                          & {\color[HTML]{000000} 80.92}                         & \cellcolor[HTML]{CFCFCF}{\color[HTML]{000000} 94.02} \\ \cline{2-6} 
{\color[HTML]{000000} }                             & {\color[HTML]{000000} gmm}  & {\color[HTML]{000000} mfcc}  & {\color[HTML]{000000} 76.48}                         & {\color[HTML]{000000} 79.39}                         & {\color[HTML]{000000} 68.82}                         \\ \cline{2-6} 
{\color[HTML]{000000} }                             & {\color[HTML]{000000} gmm}  & {\color[HTML]{000000} spect} & {\color[HTML]{000000} 86.81}                         & {\color[HTML]{000000} 81.49}                         & {\color[HTML]{000000} 86.52}                         \\ \cline{2-6} 
{\color[HTML]{000000} }                             & {\color[HTML]{000000} lstm} & {\color[HTML]{000000} spect} & {\color[HTML]{000000} 92.93}                         & {\color[HTML]{000000} 88.89}                         & {\color[HTML]{000000} 92.54}                         \\ \cline{2-6} 
{\color[HTML]{000000} }                             & {\color[HTML]{000000} rf}   & {\color[HTML]{000000} cepst} & {\color[HTML]{000000} 93.92}                         & \cellcolor[HTML]{CFCFCF}{\color[HTML]{000000} 91.98} & {\color[HTML]{000000} 93.3}                          \\ \cline{2-6} 
{\color[HTML]{000000} }                             & {\color[HTML]{000000} rf}   & {\color[HTML]{000000} mfcc}  & {\color[HTML]{000000} 94.87}                         & {\color[HTML]{000000} 90.08}                         & {\color[HTML]{000000} 93.26}                         \\ \cline{2-6} 
{\color[HTML]{000000} }                             & {\color[HTML]{000000} rf}   & {\color[HTML]{000000} spect} & {\color[HTML]{000000} 92.79}                         & {\color[HTML]{000000} 85.5}                          & {\color[HTML]{000000} 93.26}                         \\ \cline{2-6} 
{\color[HTML]{000000} }                             & {\color[HTML]{000000} svm}  & {\color[HTML]{000000} cepst} & {\color[HTML]{000000} 95.15}                         & {\color[HTML]{000000} 77.67}                         & \cellcolor[HTML]{CFCFCF}{\color[HTML]{000000} 94.57} \\ \cline{2-6} 
{\color[HTML]{000000} }                             & {\color[HTML]{000000} svm}  & {\color[HTML]{000000} mfcc}  & \cellcolor[HTML]{CFCFCF}{\color[HTML]{000000} 96.21} & {\color[HTML]{000000} 83.97}                         & \cellcolor[HTML]{CFCFCF}{\color[HTML]{000000} 96.23} \\ \cline{2-6} 
\multirow{-14}{*}{{\color[HTML]{000000} Non-mixed}} & {\color[HTML]{000000} svm}  & {\color[HTML]{000000} spect} & {\color[HTML]{000000} 75.32}                         & {\color[HTML]{000000} 69.85}                         & {\color[HTML]{000000} 77.09}                         \\ \hline
\end{tabular}
}
\end{table}

\begin{table*}
\centering
\caption{F1 Score summarization}
\label{table:results:f1}
\resizebox{\textwidth}{!}{
\begin{tabular}{@{}|l|l|l||lll||lll||lll|@{}}
\toprule
                             &      &       & \multicolumn{3}{l|}{Drug}                                                                                                                                      & \multicolumn{3}{l|}{Exhale}                                                                                                                                    & \multicolumn{3}{l|}{Inhale}                                                                                                                                    \\ \midrule \midrule
                             &      &       & \multicolumn{1}{l|}{LOSO}                                                 & \multicolumn{1}{l|}{Multi}                                                & Single & \multicolumn{1}{l|}{LOSO}                                                 & \multicolumn{1}{l|}{Multi}                                                & Single & \multicolumn{1}{l|}{LOSO}                                                 & \multicolumn{1}{l|}{Multi}                                                & Single \\ \midrule
                             & ada  & cepst & \multicolumn{1}{l|}{17.58}                                                & \multicolumn{1}{l|}{64.84}                                                & 63.28  & \multicolumn{1}{l|}{\cellcolor[HTML]{CFCFCF}{\color[HTML]{000000} 83.57}} & \multicolumn{1}{l|}{\cellcolor[HTML]{CFCFCF}{\color[HTML]{000000} 87.27}} & 84.45  & \multicolumn{1}{l|}{86.6}                                                 & \multicolumn{1}{l|}{89.24}                                                & 86.79  \\ \cmidrule(l){2-12} 
                             & ada  & mfcc  & \multicolumn{1}{l|}{11.87}                                                & \multicolumn{1}{l|}{64.1}                                                 & 65.51  & \multicolumn{1}{l|}{82.46}                                                & \multicolumn{1}{l|}{86.67}                                                & 81.03  & \multicolumn{1}{l|}{80.55}                                                & \multicolumn{1}{l|}{89.46}                                                & 85.42  \\ \cmidrule(l){2-12} 
                             & ada  & spect & \multicolumn{1}{l|}{\cellcolor[HTML]{CFCFCF}{\color[HTML]{000000} 25.89}} & \multicolumn{1}{l|}{\cellcolor[HTML]{CFCFCF}{\color[HTML]{000000} 74.85}} & 74.5   & \multicolumn{1}{l|}{72.77}                                                & \multicolumn{1}{l|}{81.48}                                                & 78.43  & \multicolumn{1}{l|}{84.73}                                                & \multicolumn{1}{l|}{\cellcolor[HTML]{CFCFCF}{\color[HTML]{000000} 91.19}} & 88.94  \\ \cmidrule(l){2-12} 
                             & cnn  & time  & \multicolumn{1}{l|}{21.85}                                                & \multicolumn{1}{l|}{58.1}                                                 & 63.41  & \multicolumn{1}{l|}{\cellcolor[HTML]{CFCFCF}{\color[HTML]{000000} 89.08}} & \multicolumn{1}{l|}{86.81}                                                & 83.06  & \multicolumn{1}{l|}{\cellcolor[HTML]{CFCFCF}{\color[HTML]{000000} 94.94}} & \multicolumn{1}{l|}{89.95}                                                & 89.05  \\ \cmidrule(l){2-12} 
                             & gmm  & cepst & \multicolumn{1}{l|}{25.81}                                                & \multicolumn{1}{l|}{63.96}                                                & 67.46  & \multicolumn{1}{l|}{72.42}                                                & \multicolumn{1}{l|}{83.73}                                                & 81.02  & \multicolumn{1}{l|}{48.3}                                                 & \multicolumn{1}{l|}{84.74}                                                & 85.21  \\ \cmidrule(l){2-12} 
                             & gmm  & mfcc  & \multicolumn{1}{l|}{}                                                     & \multicolumn{1}{l|}{17.13}                                                & 0.47   & \multicolumn{1}{l|}{72.12}                                                & \multicolumn{1}{l|}{73.85}                                                & 65.76  & \multicolumn{1}{l|}{41.62}                                                & \multicolumn{1}{l|}{53.86}                                                & 53.32  \\ \cmidrule(l){2-12} 
                             & gmm  & spect & \multicolumn{1}{l|}{\cellcolor[HTML]{CFCFCF}{\color[HTML]{000000} 32.78}} & \multicolumn{1}{l|}{\cellcolor[HTML]{CFCFCF}{\color[HTML]{000000} 73.59}} & 74.27  & \multicolumn{1}{l|}{67.42}                                                & \multicolumn{1}{l|}{72.78}                                                & 74.44  & \multicolumn{1}{l|}{\cellcolor[HTML]{CFCFCF}{\color[HTML]{000000} 89.34}} & \multicolumn{1}{l|}{89.13}                                                & 85.89  \\ \cmidrule(l){2-12} 
                             & lstm & spect & \multicolumn{1}{l|}{23.1}                                                 & \multicolumn{1}{l|}{66.61}                                                & 70.67  & \multicolumn{1}{l|}{75.5}                                                 & \multicolumn{1}{l|}{86.4}                                                 & 83.21  & \multicolumn{1}{l|}{86.04}                                                & \multicolumn{1}{l|}{90}                                                   & 88.87  \\ \cmidrule(l){2-12} 
                             & rf   & cepst & \multicolumn{1}{l|}{25.41}                                                & \multicolumn{1}{l|}{63.22}                                                & 64.73  & \multicolumn{1}{l|}{\cellcolor[HTML]{CFCFCF}{\color[HTML]{000000} 83.01}} & \multicolumn{1}{l|}{86.55}                                                & 84.77  & \multicolumn{1}{l|}{86.6}                                                 & \multicolumn{1}{l|}{89.19}                                                & 89.38  \\ \cmidrule(l){2-12} 
                             & rf   & mfcc  & \multicolumn{1}{l|}{15.87}                                                & \multicolumn{1}{l|}{59.78}                                                & 63.84  & \multicolumn{1}{l|}{82.57}                                                & \multicolumn{1}{l|}{\cellcolor[HTML]{CFCFCF}{\color[HTML]{000000} 86.97}} & 84.88  & \multicolumn{1}{l|}{84.69}                                                & \multicolumn{1}{l|}{87.57}                                                & 87.01  \\ \cmidrule(l){2-12} 
                             & rf   & spect & \multicolumn{1}{l|}{\cellcolor[HTML]{CFCFCF}{\color[HTML]{000000} 30.98}} & \multicolumn{1}{l|}{\cellcolor[HTML]{CFCFCF}{\color[HTML]{000000} 73.69}} & 75.91  & \multicolumn{1}{l|}{74.26}                                                & \multicolumn{1}{l|}{80.86}                                                & 80.83  & \multicolumn{1}{l|}{88.41}                                                & \multicolumn{1}{l|}{\cellcolor[HTML]{CFCFCF}{\color[HTML]{000000} 90.65}} & 90.27  \\ \cmidrule(l){2-12} 
                             & svm  & cepst & \multicolumn{1}{l|}{11.42}                                                & \multicolumn{1}{l|}{62.82}                                                & 65.09  & \multicolumn{1}{l|}{68.1}                                                 & \multicolumn{1}{l|}{86.9}                                                 & 84.8   & \multicolumn{1}{l|}{68.26}                                                & \multicolumn{1}{l|}{88.92}                                                & 88.77  \\ \cmidrule(l){2-12} 
                             & svm  & mfcc  & \multicolumn{1}{l|}{12.83}                                                & \multicolumn{1}{l|}{59.93}                                                & 66.03  & \multicolumn{1}{l|}{79.33}                                                & \multicolumn{1}{l|}{\cellcolor[HTML]{CFCFCF}{\color[HTML]{000000} 88.91}} & 87.66  & \multicolumn{1}{l|}{80.38}                                                & \multicolumn{1}{l|}{88.93}                                                & 88.66  \\ \cmidrule(l){2-12} 
\multirow{-14}{*}{Mixed}     & svm  & spect & \multicolumn{1}{l|}{25.59}                                                & \multicolumn{1}{l|}{71.28}                                                & 73.71  & \multicolumn{1}{l|}{40.93}                                                & \multicolumn{1}{l|}{37.5}                                                 & 35.19  & \multicolumn{1}{l|}{\cellcolor[HTML]{CFCFCF}{\color[HTML]{000000} 89.51}} & \multicolumn{1}{l|}{\cellcolor[HTML]{CFCFCF}{\color[HTML]{000000} 90.98}} & 89.73  \\ \midrule
                             & ada  & cepst & \multicolumn{1}{l|}{63.64}                                                & \multicolumn{1}{l|}{93.02}                                                & 92.93  & \multicolumn{1}{l|}{\cellcolor[HTML]{CFCFCF}{\color[HTML]{000000} 94.25}} & \multicolumn{1}{l|}{95.98}                                                & 93.18  & \multicolumn{1}{l|}{\cellcolor[HTML]{CFCFCF}{\color[HTML]{000000} 96.17}} & \multicolumn{1}{l|}{\cellcolor[HTML]{CFCFCF}{\color[HTML]{000000} 96.41}} & 93.77  \\ \cmidrule(l){2-12} 
                             & ada  & mfcc  & \multicolumn{1}{l|}{40}                                                   & \multicolumn{1}{l|}{\cellcolor[HTML]{CFCFCF}{\color[HTML]{000000} 96.43}} & 94.77  & \multicolumn{1}{l|}{91.4}                                                 & \multicolumn{1}{l|}{\cellcolor[HTML]{CFCFCF}{\color[HTML]{000000} 96.15}} & 94.09  & \multicolumn{1}{l|}{84.76}                                                & \multicolumn{1}{l|}{\cellcolor[HTML]{CFCFCF}{\color[HTML]{000000} 96.72}} & 94.03  \\ \cmidrule(l){2-12} 
                             & ada  & spect & \multicolumn{1}{l|}{56}                                                   & \multicolumn{1}{l|}{95.38}                                                & 94.91  & \multicolumn{1}{l|}{85.98}                                                & \multicolumn{1}{l|}{93.89}                                                & 92.7   & \multicolumn{1}{l|}{90.75}                                                & \multicolumn{1}{l|}{95.79}                                                & 93.43  \\ \cmidrule(l){2-12} 
                             & cnn  & time  & \multicolumn{1}{l|}{70}                                                   & \multicolumn{1}{l|}{82.8}                                                 & 85.84  & \multicolumn{1}{l|}{\cellcolor[HTML]{CFCFCF}{\color[HTML]{000000} 95.23}} & \multicolumn{1}{l|}{\cellcolor[HTML]{CFCFCF}{\color[HTML]{000000} 96.3}}  & 93.58  & \multicolumn{1}{l|}{\cellcolor[HTML]{CFCFCF}{\color[HTML]{000000} 98.24}} & \multicolumn{1}{l|}{\cellcolor[HTML]{CFCFCF}{\color[HTML]{000000} 96.77}} & 96.2   \\ \cmidrule(l){2-12} 
                             & gmm  & cepst & \multicolumn{1}{l|}{62.5}                                                 & \multicolumn{1}{l|}{93.16}                                                & 94.85  & \multicolumn{1}{l|}{88.37}                                                & \multicolumn{1}{l|}{94.92}                                                & 94.26  & \multicolumn{1}{l|}{74.23}                                                & \multicolumn{1}{l|}{96.38}                                                & 96     \\ \cmidrule(l){2-12} 
                             & gmm  & mfcc  & \multicolumn{1}{l|}{}                                                     & \multicolumn{1}{l|}{10.78}                                                & 18.43  & \multicolumn{1}{l|}{81.72}                                                & \multicolumn{1}{l|}{83.02}                                                & 76.11  & \multicolumn{1}{l|}{48.1}                                                 & \multicolumn{1}{l|}{66.32}                                                & 75.89  \\ \cmidrule(l){2-12} 
                             & gmm  & spect & \multicolumn{1}{l|}{47.06}                                                & \multicolumn{1}{l|}{94.36}                                                & 95.14  & \multicolumn{1}{l|}{81.6}                                                 & \multicolumn{1}{l|}{85.48}                                                & 83.73  & \multicolumn{1}{l|}{92.95}                                                & \multicolumn{1}{l|}{93.13}                                                & 88.58  \\ \cmidrule(l){2-12} 
                             & lstm & spect & \multicolumn{1}{l|}{69.8}                                                 & \multicolumn{1}{l|}{93.24}                                                & 92.97  & \multicolumn{1}{l|}{89.9}                                                 & \multicolumn{1}{l|}{93.55}                                                & 93.19  & \multicolumn{1}{l|}{92.93}                                                & \multicolumn{1}{l|}{94.08}                                                & 94.03  \\ \cmidrule(l){2-12} 
                             & rf   & cepst & \multicolumn{1}{l|}{\cellcolor[HTML]{CFCFCF}{\color[HTML]{000000} 84.21}} & \multicolumn{1}{l|}{92.35}                                                & 92.9   & \multicolumn{1}{l|}{\cellcolor[HTML]{CFCFCF}{\color[HTML]{000000} 91.61}} & \multicolumn{1}{l|}{94.57}                                                & 94.24  & \multicolumn{1}{l|}{\cellcolor[HTML]{CFCFCF}{\color[HTML]{000000} 97.07}} & \multicolumn{1}{l|}{96.1}                                                 & 94.47  \\ \cmidrule(l){2-12} 
                             & rf   & mfcc  & \multicolumn{1}{l|}{\cellcolor[HTML]{CFCFCF}{\color[HTML]{000000} 74.07}} & \multicolumn{1}{l|}{93.61}                                                & 93.8   & \multicolumn{1}{l|}{89.95}                                                & \multicolumn{1}{l|}{95.68}                                                & 93.58  & \multicolumn{1}{l|}{95.24}                                                & \multicolumn{1}{l|}{95.48}                                                & 95.26  \\ \cmidrule(l){2-12} 
                             & rf   & spect & \multicolumn{1}{l|}{66.67}                                                & \multicolumn{1}{l|}{94.9}                                                 & 95.72  & \multicolumn{1}{l|}{86.03}                                                & \multicolumn{1}{l|}{92.57}                                                & 92.73  & \multicolumn{1}{l|}{93.51}                                                & \multicolumn{1}{l|}{95.15}                                                & 94.47  \\ \cmidrule(l){2-12} 
                             & svm  & cepst & \multicolumn{1}{l|}{43.9}                                                 & \multicolumn{1}{l|}{93.81}                                                & 94.59  & \multicolumn{1}{l|}{78.28}                                                & \multicolumn{1}{l|}{96.05}                                                & 95.3   & \multicolumn{1}{l|}{83.25}                                                & \multicolumn{1}{l|}{95.79}                                                & 96.04  \\ \cmidrule(l){2-12} 
                             & svm  & mfcc  & \multicolumn{1}{l|}{47.62}                                                & \multicolumn{1}{l|}{\cellcolor[HTML]{CFCFCF}{\color[HTML]{000000} 96.18}} & 96.76  & \multicolumn{1}{l|}{86.85}                                                & \multicolumn{1}{l|}{\cellcolor[HTML]{CFCFCF}{\color[HTML]{000000} 96.54}} & 96.62  & \multicolumn{1}{l|}{85.31}                                                & \multicolumn{1}{l|}{96.24}                                                & 95.76  \\ \cmidrule(l){2-12} 
\multirow{-14}{*}{Non-mixed} & svm  & spect & \multicolumn{1}{l|}{\cellcolor[HTML]{CFCFCF}{\color[HTML]{000000} 81.82}} & \multicolumn{1}{l|}{\cellcolor[HTML]{CFCFCF}{\color[HTML]{000000} 95.96}} & 96.81  & \multicolumn{1}{l|}{75.37}                                                & \multicolumn{1}{l|}{76.28}                                                & 75.89  & \multicolumn{1}{l|}{95.36}                                                & \multicolumn{1}{l|}{94.98}                                                & 94.24  \\ \bottomrule
\end{tabular}
}
\end{table*}

\subsection{Key performance indicators}

Table~\ref{table:results:Accuracy} summarizes the classification accuracy for drug actuation, exhalation and inhalation sounds, across all validation setups. Out of a superficial examination, no method is better, but key performance indicators need to be established. Multi-subject and single-subject settings indicate the classifier's success, if the user has previously participated in the sampling process. Leave-one-subject-out setting is the closest to a real situation setting, since a commercial classifier would not assume that a new user has already submitted samples to the training process. Even though a feedback loop can improve accuracy \cite{nousias2018mhealth}, such a process cannot be a prerequisite. As a result, leave-one-subject-out performance is the most representative. Furthermore, between mixed and non-mixed setup, the first is closer to the real situation, since the non-mixed assumes that the position of audio segments containing a certain audio event, is known before the classifier is applied. In short, the key performance indicators can be summarized below:
\begin{itemize}
    \item The three highest non-mixed LOSO accuracies should be highlighted (Table~\ref{table:results:Accuracy})
    \item The three highest mixed multi-subject accuracies should be highlighted (Table~\ref{table:results:Accuracy})
    \item The three highest mixed LOSO accuracies should be highlighted (Table~\ref{table:results:Accuracy})
    \item The three highest drug precision mixed and non-mixed sensitivity should be highlighted (Table~\ref{table:results:Accuracy})
\end{itemize}
Metrics to measure the performance of the compared classifiers are accuracy, sensitivity and specificity. For the sake of self completeness:

\begin{equation}
    \text{accuracy}=\frac{(TP+TN)}{(TP+FP+TN+FN)}
\end{equation}

\begin{equation}
    \text{precision}=\frac{(TP)}{(TP+FP)}
\end{equation}

\begin{equation}
   \text{recall} =\frac{TN}{(TN+FP)}
\end{equation}

where $TP$ is the number of positive correct identifications, $TN$ the number of negative correct identifications, $FP$ is the number of positive incorrect identifications and $FN$ is the number of negative incorrect identifications.

\subsection{Results}

Close inspection of tables ~\ref{table:results:Accuracy} and ~\ref{table:results:f1} reveals that no method clearly outperforms the others. The performance of data-driven approaches largely depends on the dataset and the pre-processing steps. However, since all the approaches were trained with the same dataset, we can quantify which can capture the individual characteristics more efficiently. For all methodologies, the multi-subject approach yields the highest score. This means that if user's data have been included in the training process, the success probability climbs to a level near 96\%. 

Given the Key Performance Indicators (KPIs) defined in the previous section, in the non-mixed setup, SVM-MFCC, ADA-MFCC and CNN-TIME yield the highest accuracy, when patients' data are already included in the dataset, corresponding to Multi and Single subject cases. However, in the "Leave One Subject Out" case, which also corresponds to a real-life scenario CNN-TIME, ADA-CEPST, and RF-CEPST yield the best results. In the mixed setup, the results are similar with LSTM-SPECT to replace CNN-TIME only in the multi-subject setup. Further insight is provided by table~\ref{table:results:f1}. Measuring F1 score is highly important, since it consists of the harmonic mean of precision and recall of a candidate detector. A close inspection reveals that drug detection sounds have much more different characteristics than exhalation and inhalation signals. For the Exhale/LOSO case, CNN-TIME has the best performance that, also, concurs with the results presented in the Accuracy Table. For detecting inhalations in LOSO setup, CNN-TIME also yields the best performance. However, in the multi-subject setup, MFCC methods demonstrate a more stable performance than the CNN-TIME method. In the drug classification case, the F1 score reveals much more different results. Specifically, the mixed setup demonstrates lower F1 Scores than the non-mixed setup, meaning that "Drug" detection is highly affected by the surrounding audio events. ADA-SPECT, GMM-SPECT and RF-SPECT have the highest accuracy in multi-subject mixed setup, reaching over 73\%, while for the non-mixed case, the accuracy is over 96\%. Likewise, LOSO with Non-mixed on Random Forest based methods reaches 84\%, but in the mixed setup the corresponding performance is 32\%.

\begin{figure}[H]
\centering
 \includegraphics[width=\linewidth]{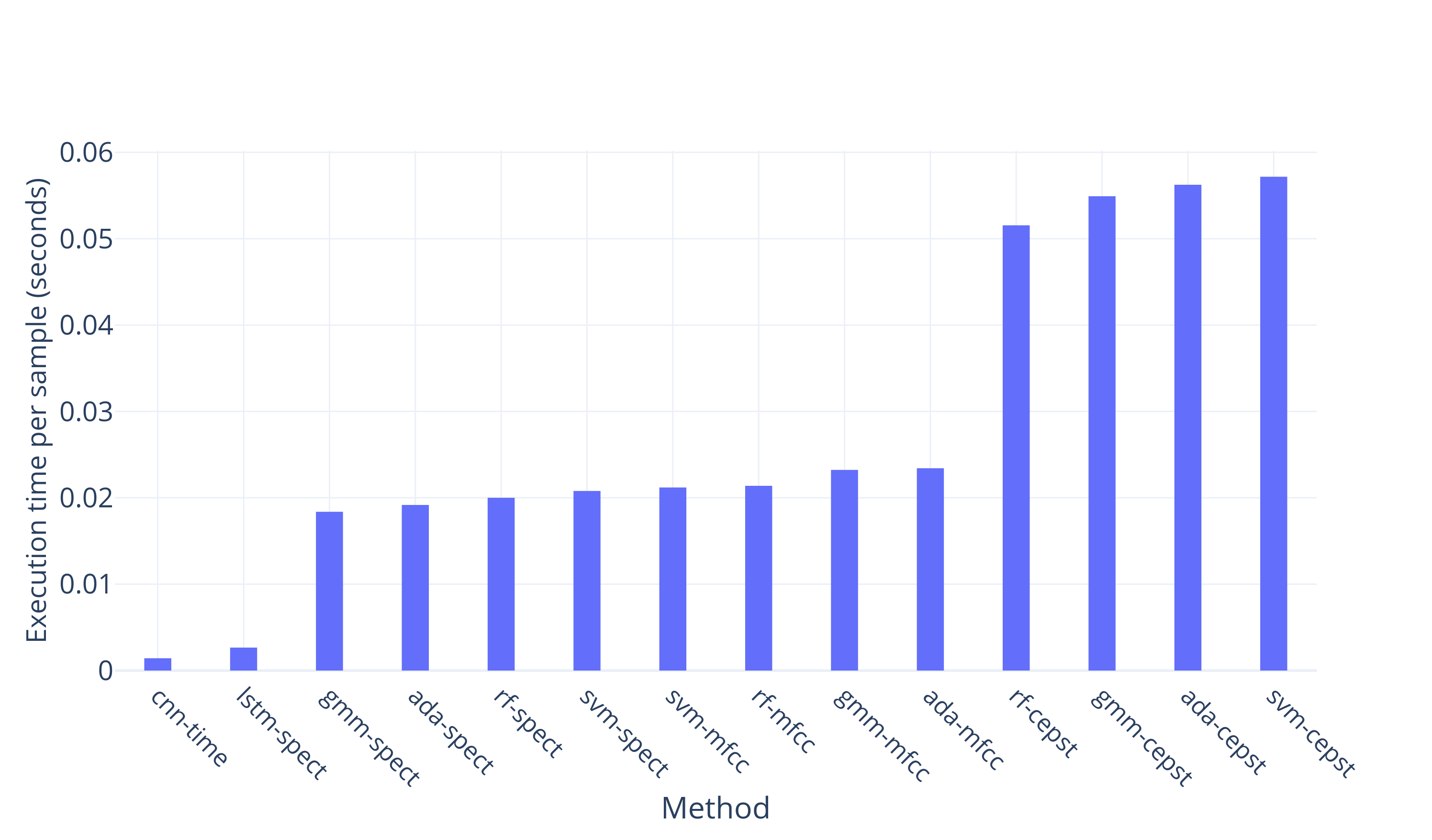}
\caption{Comparison of the computational cost of CNN based approach and other studies.}
\label{fig:execution-time}  
\end{figure}

Finally, we compare the computational cost of the CNN model with the other approaches, executed in the same machine (Intel(R) Core(TM) i5-5250U CPU @ 2.7GHz). The results are summarized in figure \ref{fig:execution-time}. This figure highlights the gain in computational speedup, compared to the time-consuming spectrogram and similar feature-based algorithms. Specifically, figure \ref{fig:execution-time} shows that classification by CNN is 40 times faster than the slowest cepstrogram-based methods and fifteen times faster than spectrogram and MFCC-based methods.

\begin{table}
\centering
\caption{Comparison of the computational cost of CNN based approach and other studies (measured in milliseconds)}
\label{table:result:performance_comparison}
\resizebox{\linewidth}{!}{
\begin{tabular}{@{}|l|l|l|l|@{}}
\toprule
Method     & Classification execution time & Feature extraction Time & Sum    \\ \midrule
cnn-time   & 0.000460227                   & 0.000968868             & 0.0014 \\ \midrule
lstm-spect & 0.000700771                   & 0.001955032             & 0.0027 \\ \midrule
gmm-spect  & 1.82257E-05                   & 0.018367229             & 0.0184 \\ \midrule
ada-spect  & 0.000773556                   & 0.018398599             & 0.0192 \\ \midrule
rf-spect   & 0.000278274                   & 0.019732354             & 0.02   \\ \midrule
svm-spect  & 4.70897E-05                   & 0.02074642              & 0.0208 \\ \midrule
svm-mfcc   & 2.1926E-05                    & 0.021171822             & 0.0212 \\ \midrule
rf-mfcc    & 0.000275223                   & 0.021114178             & 0.0214 \\ \midrule
gmm-mfcc   & 2.35955E-05                   & 0.023210378             & 0.0232 \\ \midrule
ada-mfcc   & 0.000852494                   & 0.022573248             & 0.0234 \\ \midrule
rf-cepst   & 0.000272068                   & 0.051282082             & 0.0516 \\ \midrule
gmm-cepst  & 1.19811E-05                   & 0.054902331             & 0.0549 \\ \midrule
ada-cepst  & 0.0008844                     & 0.055359981             & 0.0562 \\ \midrule
svm-cepst  & 3.10628E-05                   & 0.057136457             & 0.0572 \\ \bottomrule
\end{tabular}
}
\end{table}

\section{Discussion and Future Considerations}
\label{section:future}

The pressurized metered dose inhaler (pMDI) is the most commonly used inhaler, with total worldwide sales of pMDI products reaching over \$2 billion per year \cite{horne2005concordance}. Studies have reported that over 50\% of patients are prone to not adhering to the correct inhaler technique \cite{bender1997nonadherence}. It has been stated that acoustics can be employed to detect and recognize dry powder inhaler sounds \cite{murphy2012relationship}. Almost 4 decades ago, the earliest investigations into poor inhaler sufficiency in patients using pMDIs, indicated that poor technique was likely to be associated with less than the highest quality of response to the therapy \cite{oprehek1976patient, paterson1976use}. Several studies have since intimated a strong connection between poor inhaler adequacy and patient outcomes, on a larger scale \cite{e2001determinants}, including higher rates of hospital access and emergency room attendance \cite{melani17}. By providing suitable feedback to medical staff and guiding patients to improve their inhaler usage technique, we could facilitate efficient self-management of obstructive respiratory diseases, allowing patients to avoid dangerous exacerbation events \cite{wu2015max}. Asthma is almost in the center of the wave of digital health developments, as it requires systematic attention of both health care specialist doctors and patients \cite{dhanani2016fundamentals}. Employing acoustic signal processing methods, the aforementioned algorithms were developed to accurately identify drug actuations, from pMDI's. As science and technology evolve and modern sensing components are becoming more available, a continual improvement process of inhalers with an extended range of monitoring capabilities, holds the promise to further optimize asthma self-management. These methods provide an opportunity to enhance clinical education, by providing informative feedback to patients, which may contribute to improving respiratory health. Future work will consist of identifying pMDI inhalations to monitor actuation coordination technique and to provide patient feedback, regarding drug delivery using acoustic methods. The accurate estimation of the parameters would be of significant clinical benefit to both patients and healthcare professionals, by enhancing precision medicine for chronic respiratory diseases. Ultimately, this work aims to prevent common mistakes, leading to potential upcoming dangerous events, such as exacerbation and hospitalization.

\section{Conclusions}
\label{section:conclusions}

Asthma forms an important socioeconomic burden, both in terms of medication costs and disability adjusted life years. The accurate assessment of the state of asthma is the fundamental basis of digital health approaches and also is the most significant factor towards the preventive and efficient management of the disease. The necessity of inhaled medication offers a basic platform, upon which, modern technologies can be integrated, namely the inhaler device system itself. The control of asthma is a complex and multiparametric issue, that is greatly affected not only by physiological and environmental parameters, but, also, the psychological state of patients and their cultural and socioeconomic background. Indicative of the complexity of asthma disease, is the diversity of its prevalence around the world. All the above outline the need to increase the active involvement of patients, in modern treatment methodologies and to use modern technologies so as to create easy-to-use tools, for safe and effective self-management. A fundamental step in this direction is the creation of a sensing framework, that could provide accurate information, about the health of patients and help their doctors understand any possible difficulty, that prevents patients from using their inhaled medication adherence correctly. This need for the modernization of inhaler devices, has stimulated the research and commercial interest for their enhancement with novel sensing capabilities and has led to a number of approaches, that focus mainly on the detection of inhaler actuations. The modern adherence monitoring environment has also been analyzed in other studies, addressing important related issues, such as the interpretation of results and the design of interventions that promote adherence.

\appendices

\section{Machine learning algorithms}

\subsection{Support vector machines}
\label{appendix:SVM}

For a decision hyper-plane ${\bf x}^T {\bf w}+b=0$ to separate the two classes: $P=\{({\bf x}_i,1)\}$ and $N=\{({\bf x}_i,-1)\}$, it has to satisfy
\[
y_i ({\bf x}_i^T{\bf w}+b) \ge 0	
\]
for both ${\bf x}_i \in P$ and ${\bf x}_i \in N$. Among all such planes satisfying this condition, we want to find the optimal one $H_0$ that separates the two classes with the maximal margin (the distance between the decision plane and the closest sample points). The optimal plane should be in the middle of the two classes, so that the distance from the plane to the closest point on either side is the same. We define two additional planes $H_+$ and $H_-$ that are parallel to $H_0$ and go through the point closest to the plane on either side:
\[
  {\bf x}^T {\bf w}+b=1,\;\;\;\;\mbox{and}\;\;\;\;{\bf x}^T {\bf w}+b=-1	
\]
All points ${\bf x}_i \in P$ on the positive side should satisfy:
\[
  {\bf x}_i^T {\bf w}+b \ge 1,\;\;\;\;y_i=1	
\]
and all points ${\bf x}_i \in N$ on the negative side should satisfy:
\[
  {\bf x}_i^T {\bf w}+b \le -1,\;\;\;\; y_i=-1	
\]
These can be combined into one inequality:
\[
y_i ({\bf x}_i^T{\bf w} +b) \ge 1,\;\;\;(i=1,\cdots,m)	
\]
The equality holds for those points on the planes $H_+$ or $H_-$. Such points are called {\em support vectors}, for which
\[
  {\bf x}_i^T {\bf w}+b = y_i	
\]
i.e., the following holds for all support vectors:
\[
b=y_i-{\bf x}_i^T {\bf w}=y_i-\sum_{j=1}^m  \alpha_j y_j ({\bf x}_i^T {\bf x}_j)	
\]
Moreover, the distances from the origin to the three parallel planes $H_-$, $H_0$ and $H_+$ are, respectively: $|b-1|/||{\bf w}||$, $|b|/||{\bf w}||$
and $|b+1|/||{\bf w}||$ and the distance between planes $H_-$ and $H_+$ is $2/||{\bf w}||$. Our goal is to maximize this distance, or equivalently, to minimize the norm $||{\bf w}||$. Now the problem of finding the optimal decision plane in terms of ${\bf w}$ and $b$ can be formulated as:
\begin{eqnarray}
  &	\mbox{minimize} & \frac{1}{2}{\bf w}^T {\bf w}=\frac{1}{2}||{\bf w}||^2
  \;\;\;\;\;\;\mbox{(objective function)}	\nonumber \\
  &	\mbox{subject to} & y_i ({\bf x}_i^T {\bf w}+b) \ge 1,\;\;\mbox{or}\;\;
  1-y_i ({\bf x}_i^T {\bf w}+b) \le 0,
  \nonumber
\end{eqnarray}
$(\mathrm{i}=1, \cdots, m)$. Since the objective function is quadratic, this constrained optimization problem is called a quadratic program (QP) problem. If the objective function is linear instead, the problem is a linear program (LP) problem. This QP problem can be solved by Lagrange multipliers method to minimize the following
\[
L_p({\bf w},b,{\bf \alpha})=\frac{1}{2}||{\bf w}||^2
+\sum_{i=1}^m \alpha_i(1-y_i({\bf x}_i^T{\bf w}+b))
\]
with respect to ${\bf w}$, $b$ and the Lagrange coefficients $\alpha_i\ge 0$, where $(i=1,\cdots,\alpha_m)$. We let
\[ 	\frac{\partial}{\partial w}L_p({\bf w},b)=0,\;\;\;
	\frac{\partial}{\partial b}L_p({\bf w},b)=0	\]
These lead, respectively, to
\[	{\bf w}=\sum_{j=1}^m \alpha_j y_j {\bf x}_j,\;\;\;\mbox{and}\;\;\;\;
	\sum_{i=1}^m \alpha_i y_i=0	\]
Substituting these two equations back into the expression of $L({\bf w},b)$, we get the {\em dual problem} (with respect to $\alpha_i$) of the above
{\em primal problem}:
\begin{eqnarray}
& 	\mbox{maximize} & L_d({\bf \alpha})=
	\sum_{i=1}^m\alpha_i -\frac{1}{2}
	\sum_{i=1}^m \sum_{j=1}^m \alpha_i \alpha_j y_i y_j {\bf x}_i^T,{\bf x}_j
	\nonumber \\
&	\mbox{subject to} & \alpha_i\ge 0,\;\;\;\;
	\sum_{i=1}^m \alpha_i y_i=0	\nonumber
\end{eqnarray}
The dual problem is related to the primal problem by:
\[	L_d({\bf \alpha})=inf_{({\bf w},b)} L_p({\bf w},b,{\bf \alpha})	\]
i.e., $L_d$ is the greatest lower bound (infimum) of $L_p$ for all ${\bf w}$ and $b$. Solving this dual problem (an easier problem than the primal one), we get $\alpha_i$, from which ${\bf w}$ of the optimal plane can be found. Those points ${\bf x}_i$ on either of the two planes $H_+$ and $H_-$ (for which the equality $y_i({\bf w}^T {\bf x}_i+b)=1$ holds) are called {\em support vectors} and they correspond to positive Lagrange multipliers $\alpha_i>0$. The training depends only on the support vectors, while all other samples away from the planes $H_+$ and $H_-$, are not important. For a support vector ${\bf x}_i$ (on the $H_-$ or $H_+$ plane), the constraining condition is
\[	y_i \left({\bf x}_i^T {\bf w}+b\right) = 1\;\;\;\;(i \in sv)	\]
here $sv$ is a set of all indices of support vectors ${\bf x}_i$ (corresponding to $\alpha_i > 0$). Substituting
\[	{\bf w}=\sum_{j=1}^m \alpha_j y_j {\bf x}_j=\sum_{j\in sv} \alpha_j y_j {\bf x}_j \]
we get
\[	y_i(\sum_{j\in sv} \alpha_j y_j {\bf x}^T_i {\bf x}_j+b) = 1	\]
Note that the summation only contains terms corresponding to those support vectors ${\bf x}_j$, with $\alpha_j>0$, i.e.
\[	y_i \sum_{j\in sv} \alpha_j y_j {\bf x}^T_i {\bf x}_j= 1-y_i b	\]
For the optimal weight vector ${\bf w}$ and optimal $b$, we have:
\begin{eqnarray}
||{\bf w}||^2&=&{\bf w}^T {\bf w} = \sum_{i\in sv} \alpha_i y_i {\bf x}^T_i
	\sum_{j\in sv} \alpha_j y_j {\bf x}_j	\nonumber \\
&=&	\sum_{i\in sv} \alpha_i (1-y_i b)
	=\sum_{i\in sv} \alpha_i - b\sum_{i\in sv} \alpha_i y_i	\nonumber \\
&=&	\sum_{i\in sv} \alpha_i \nonumber
\end{eqnarray}
The last equality is due to $\sum_{i=1}^m \alpha_i y_i=0$ shown above. Recall that the distance between the two margin planes $H_+$ and $H_-$ and the margin is $2/||{\bf w}||$, the distance between $H_+$ (or $H_-$) and the optimal decision plane $H_0$ is
\[	\frac{1}{||{\bf w}||}=\left(\sum_{i\in sv} \alpha_i\right)^{-1/2}	\]

\subsection{Adaboost}
\label{appendix:Adaboost}

Let's assume $N$ samples $x_{i} \in \mathrm{X}$, $i=\{1,...,N\}$ with corresponding labels $y_{i} \in Y=\{-1,+1\}$ are given and used as a training set
$\left(x_{1}, y_{1}\right),\left(x_{2}, y_{2}\right), \ldots \ldots\left(x_{n}, y_{N}\right)$. 

Initially set up a weight $\mathrm{D}(\mathrm{i})$ and make $\mathrm{D}(\mathrm{i})=1 / \mathrm{N}$ and then iterate for $\mathrm{t}=1,2, \ldots, \mathrm{T}$, where $\mathrm{T}$ is a parameter representing the maximum circulation times of training. For T-training, firstly, the weight distributing on the sample $\left\{X_{i}, \quad Y_{i}\right\}$ is recorded as $\mathrm{D}_{\mathrm{t}}(\mathrm{i})$ while the T-th iteration happens and as per the distribution $\mathrm{D}_{\mathrm{t}}$ the weak learner finds its weak hypothesis $\mathrm{h}_{\mathrm{t}}: \mathrm{X} \rightarrow\{+1,-1\}$ and adjusts distribution. Secondly, the error rate in computing $\mathrm{h}_{\mathrm{t}}$: $\varepsilon_{t}=\sum_{i=1}^{N} D_{t}\left(x_{i}\right)\left[h_{t}\left(x_{i}\right) \neq y_{i}\right]$. Thirdly, we compute the weight of weak classifier based on the error rate: $\alpha_{t}=(1 / 2) \ln \left(\left(1-\varepsilon_{v}\right) / \varepsilon_{t}\right)$. Fourthly, we update the sample weight: $D_{t+1}\left(x_{i}\right)=\frac{D_{t}\left(x_{i}\right)}{Z_{t}} \exp \left(-\alpha_{t} y_{i} h_{t}\left(x_{i}\right)\right)$ among which
$Z_{t}=\sum_{i=1}^{n} D_{t}\left(x_{i}\right) \exp \left(-\alpha_{t} y_{i} h_{t}\left(x_{i}\right)\right)$, where $Z_{t}$ is a standardized factor that meets the probability distribution. T-weak classifiers are gained after T times of circulation and a strong classifier
$H(x)=\operatorname{sign}\left(\sum_{t=1}^{T} \alpha_{t} h_{t}(x)\right)$ is gained after adding to the updated weight \cite{nousias2016monitoring}.

\subsection{GMM}
\label{appendix:GMM}

In the GMM classifier, the Gaussian mixture density of each d-dimensional feature vector $v$ is modeled as a linear combination of multivariate Gaussian PDFs with the general form:

\begin{equation}
    p\left(v | \theta_i\right) = \frac{1}{\left(2\pi\right)^\frac{d}{2} |C_i|^2}e^{\left[ -\frac{1}{2}\left(v - \mu_i\right)^T{C_i}^{-1}\left(v-\mu_i\right)\right]}
\end{equation}

\noindent where $\theta_i = \left(\mu_i,C_i\right)$ are the parameters of component $i$, including the mean feature vector $\mu_i$ and the $d \times d$ covariance matrix $C_i$ and $|C_i|$ is the determinant of $C_i$. The complete set of parameters for a mixture model with $K$ components, is $\Theta=\{\alpha_1,...,\alpha_K,\theta_1,...,\theta_K\}$. Each GMM model $\lambda_n$ for class $n$ is parameterized as follows:

\begin{equation}
    \lambda_n = \{{a_k}^n, {\mu_k}^n, {C_k}^n\}
\end{equation}

\noindent where $k=1,...,K$. It is important to note that each multivariate Gaussian PDF is completely defined, if we know $\theta$.

At this point, we analyze the expectation-maximization (EM) algorithm employed to compute the GMM parameters. The membership weight of data point $v$ in component $k$ given parameter $2$, is defined as:

\begin{equation}
    w_{ik } = \frac{p_k\left(v_i, \theta_k\right)a_k}{\sum_{m=1}^{K}p_m\left(v_i|\theta_m\right)a_m}
\end{equation}

for all components $k$, $1 \leq k \leq K$ and all data samples $i$, $1 \leq i \leq N$. In each iteration of the EM algorithm for Gaussian Mixtures, we deploy an E-step and an M-step. At E-Step, we compute $w_{ik}$ for all feature vectors $v_i$ and all mixture components $k$. At M-Step, we calculate the new parameters. Given $N_k = \sum_{i=1}^{N}w_{ik}$ the sum of membership weights for the $k$-th component, we get the mixture weights:

\begin{equation}
    {a_k}^{new} = \frac{N_k}{N}, \quad\quad 1 \leq k \leq K
\end{equation}

The updated mean:

\begin{equation}
{\mu_k}^{new} = \frac{1}{N_k}\sum_{i=1}^{N}w_{ik}v_i, \quad \quad 1 \leq k \leq K
\end{equation}

and the updated covariance:

\begin{equation}
    {C_k}^{new} = \frac{1}{N_k}\sum_{i=1}^{N}w_{ik}\left(v-\mu_i\right)^T\left(v-\mu_i\right)
\end{equation}

The termination criteria for the EM is the following:

\begin{equation}
    \log {l\left(\Theta\right)_{t+1}} - \log {l\left(\Theta\right)_t \leq \epsilon}
\end{equation}

\noindent where the log-likelihood, defined as $\log {l\left(\Theta\right)} = \sum_{i=1}^{N}\log {p\left(v_i|\Theta\right)}$ and $\epsilon$ is a small user-defined scalar value. In order to find the best fit for the data, we compute the GMM for $1$ to $d=40$ components iterating over full and diagonal covariance matrices, where $d$ is the size of each feature vector $v$. With the generation of each model, we estimate the Bayesian Information Criteria (BIC). The model with the lowest BIC best fits the input data.

\section{Feature extraction}

\subsection{Continuous Wavelet Transform}
\label{appendix:CWT}

In theory, it is assumed that continuous signal $x(t)$ can be approximated perfectly by $a_{n}$ and $\psi_{n}$, but in reality this is often not true. Let us consider a real signal $x(t)$ and its approximation $\hat{x}(t)$, which can be perfectly approximated by $a_{n}$ and $\psi_{n}$. The approximation error is defined as the difference between $x(t)$ and $\hat{x}(t)$. This results in an error function, which is used as a measure for the overall error. Therefore the sum of the remaining squared inner products is calculated (2-norm) and is used as a measure for the error $(\epsilon)$.

\begin{equation}
\epsilon[M]=\|x(t)-\hat{x}(t)\|^{2}=\sum_{n=M}^{+\infty}\left|\left\langle x(t), \psi_{n}(t)\right\rangle\right|^{2}
\end{equation}
with
\begin{equation}
\hat{x}(t)=\sum_{n=0}^{M-1}\left\langle x(t), \psi_{n}(t)\right\rangle \psi_{n}
\end{equation}

If the number of terms $M$ is increased the error becomes smaller i.e. when $M$ goes to $+\infty$ then the approximation error goes to zero. The rate that determines how fast $\epsilon[M]$ goes to zero, with increasing $M$ is called the decay rate and gives information about how well a certain frame can approximate a signal. To be "admissible" as a wavelet, this function must have zero mean and be localized in both time and frequency space \cite{farge1992wavelet}.

The Continuous Wavelet Transform (CWT) was introduced almost 3 decades ago, in order to overcome the limited time-frequency localization of the FFT \cite{kron} for non-stationary signals and was found to be suitable in multiple applications \cite{kron, parad}. It is similar to the human ear which exhibits similar time-frequency resolution characteristics \cite{tzan, das2}. While the Fourier Transform decomposes a signal into infinite length sines and cosines, effectively losing all time-localization information, the CWT's basis functions are scaled and shifted versions of the time-localized mother wavelet. The CWT of $x(t)$ at any scale $s$ and position $u$ is the projection of $x$ on the corresponding wavelet atom $\psi$, as described in the following formula:

\begin{equation}
W_{\psi}^{x}(u, s)=\left\langle x, \psi_{u, s}\right\rangle=\int_{-\infty}^{+\infty} x(t) \frac{1}{\sqrt{s}} \psi^{*}\left(\frac{t-u}{s}\right) d t
\end{equation}

It represents one-dimensional signals by highly redundant time-scale images in $(u, s)$. The CWT is an excellent tool for mapping the changing properties of non-stationary signals. The CWT consists of $N$ spectral values for each scale used, each of these requiring an IFFT. The computational load of the CWT and its memory requirements are thus considerable. The benefit from this high measure of redundancy in the CWT is an accurate time-frequency spectrum. Unlike a Fourier decomposition which always uses complex exponential (sine and cosine) basis functions, a wavelet decomposition uses a time-localized oscillatory function as the analyzing or mother wavelet. The mother wavelet is a function that is continuous in both time and frequency and serves as the source function, from which scaled and translated basis functions are constructed. The mother wavelet can be complex or real and it, generally, includes an adjustable parameter which controls the properties of the localized oscillation.

\end{document}